\documentclass[11pt]{article}
\usepackage{jheppubmod}
\usepackage{cite}
\usepackage{cancel}
\usepackage{xcolor}
\usepackage{caption}  
\usepackage{graphicx} 
\usepackage{float} 
\usepackage{cite}
\usepackage{relsize}
\usepackage{physics}
\usepackage{psfrag}
\usepackage{cancel}
\usepackage{array}
\usepackage{amssymb}
\usepackage{amsmath}
\usepackage{cite}
\usepackage[compat=1.1.0]{tikz-feynman}
\usepackage{amsthm}
\usepackage{float}
\usepackage{tikz}
\usetikzlibrary{decorations.markings}
\usepackage{tikz,lipsum,lmodern}
\usepackage[most]{tcolorbox}
\usepackage{hyperref}
\usepackage{xcolor}
\usepackage{comment}
\usepackage{soul}

\definecolor{linkblue}{rgb}{0.1,0.3,.7}
\definecolor{forestgreen(web)}{rgb}{0.13, 0.55, 0.13}
\definecolor{lava}{rgb}{0.81, 0.06, 0.13}
\hypersetup{
	breaklinks,
	colorlinks,
	citecolor=linkblue,
	filecolor=linkblue,
	linkcolor=linkblue,
	urlcolor=linkblue
}

\def\ee{\end{equation}}
\def\be{\begin{equation}}
\def\bea{\begin{eqnarray}}
\def\eea{\end{eqnarray}}
\newcommand{\beq}{\begin{eqnarray}}
\newcommand{\eqq}{\end{eqnarray}}
 \newcommand{\badat}{\begin{alignedat}}
 \newcommand{\eadat}{\end{alignedat}}

\newcommand{\eal}[1]{\be \begin{aligned} #1 \end{aligned}\end{equation}} 
\newcommand{\eqn}[1]{\be #1 \end{equation}} 
\newcommand{\eqa}[1]{\bea  #1\end{eqnarray}}

\long\def\new#1\endnew{{\bf #1}}		
\long\def\del#1\enddel{}

\def\del{\partial}

\usepackage{color}

\newcommand{\pink}[1]{\textcolor{\pink}{#1}}

\definecolor{dblue}{rgb}{0.2,0.50,0.80}

\def\AC#1{{\color [rgb]{0.50,0.0,0.50} [AC: #1]}}

\title{ Information Scrambling with Higher--Form Fields}

\author{Karunava Sil$^a$, Sourav Maji$^b$, Stavros Christodoulou$^c$, Abhishek Chowdhury$^b$}

\affiliation[a]{Department of Physics, New Alipore College, L Block, New Alipore, Kolkata, West Bengal, 700053, India}
\affiliation[b]{Department of Physics, School of Basic Sciences, Indian Institute of Technology Bhubaneswar,
Jatni, Khurda, Odisha, 752050, India}
\affiliation[c]{Department of Physics, University of Cyprus, Nicosia, 1678, Cyprus}

\emailAdd{karunavasil@gmail.com}
\emailAdd{sm89@iitbbs.ac.in}
\emailAdd{christodoulou.stavros@ucy.ac.cy} \emailAdd{achowdhury@iitbbs.ac.in}

\abstract{The late time behavior of $\mathrm{OTOC}$s involving generic non--conserved local operators show exponential decay in chaotic many body systems. However, it has been recently observed that for certain holographic theories, the $\mathrm{OTOC}$ involving the $U(1)$ conserved current for a gauge field instead varies diffusively at late times. The present work generalizes this observation to conserved currents corresponding to higher--form symmetries that belong to a wider class of symmetries known as generalized symmetries. We started by computing the late time behavior of $\mathrm{OTOC}$s involving $U(1)$ current operators in five dimensional AdS--Schwarzschild black hole geometry for the 2--form antisymmetric $B$--fields. The bulk solution for the $B$--field exhibits logarithmic divergences near the asymptotic AdS boundary which can be regularized by introducing a double trace deformation in the boundary CFT. Finally, we consider the more general case with antisymmetric $p$--form fields in arbitrary dimensions. In the scattering approach, the boundary $\mathrm{OTOC}$ can be written as an inner product between asymptotic `in' and `out' states which in our case is equivalent to computing the inner product between two bulk fields with and without a shockwave background.  We observe that the late time $\mathrm{OTOC}$s have power law tails which seems to be a universal feature of the higher--form fields with $U(1)$ charge conservation. 

}

\begin{document}
\maketitle

\section{Introduction}
\label{sec:intro}
A fundamental interest in recent years is to understand the mechanism of thermalization of an isolated quantum many body system. Simply speaking, thermalization in a quantum many body system is a process by virtue of which the system effectively looses the memory of its initial state as it is evolved in time generated by some complicated Hamiltonian \cite{BS}. Let us consider two generic orthogonal states $\psi_{1}$ and $\psi_{2}$ that describe two possible initial quantum states of a particular system. The orthogonality of these two states can be manifest by computing the expectation value of some simple operators $\hat{A}$ (for example the Pauli matrices) providing two completely distinct results, $\langle \psi_1|\hat{A}|\psi_1 \rangle \neq \langle \psi_2|\hat{A}|\psi_2 \rangle $. However, as time evolves, the above expectation values become nearly equal and hence it becomes impossible to distinguish the two quantum states as far as the expectation value of operator $\hat{A}$ is concerned. So, the inability to distinguish between two time evolved states using simple observables is the result of thermalization such that at late times $\langle \psi_1|\hat{A}|\psi_1\rangle= \langle \psi_2|\hat{A}|\psi_2\rangle $ and equals the corresponding expectation value in a thermal ensemble $\rho$. As the system thermalizes, it looses any initial quantum information that can be recovered by means of doing local measurements. In other words, the information is said to be spread or delocalized. This phenomenon is also known as the scrambling of quantum information \cite{Hayden_2007,Sekino:2008he}.  

For strongly coupled many body systems, any coherent excitation does not live for very long and hence scrambles quickly \cite{BS}. A measure of such fast scrambling in strongly coupled systems is given by the thermal expectation value of the double commutator involving two local operators $V$ and $W$,
\begin{equation} \label{otoceq}
C(t,x)=\left \langle [V(t,x),W(0,0)]^2 \right \rangle_{\beta}\approx \langle V(t,x)W(0,0)V(t,x)W(0,0)\rangle_{\beta},
\end{equation}
where the expectation value is evaluated in some appropriate equilibrium thermal state with inverse temperature $\beta$. The Out of Time Ordered Correlation (OTOC) function appearing in the final term of the above equation contains all the relevant information about $C(t,x)$ \cite{jahnke2019recentdevelopmentsholographicdescription}. Physically, $C(t,x)$ measures how any measurement at a spacetime point $(t,x)$ corresponding to the expectation value $\langle V\rangle$, is effected by some perturbation $W$ in the past \cite{jahnke2019recentdevelopmentsholographicdescription,Shenker:2014cwa}. The local operators $V$, $W$ in the Heisenberg picture are evolved under the unitary time evolution of the system. It has been observed that for a large class of chaotic systems, like spin--chains \cite{Chen_2016,Luitz_2017,Bohrdt_2017,Heyl_2018,Lin_2018,Xu_2019}, higher dimensional SYK--models \cite{kitaev, subir1992, Polchinski:2016xgd, Maldacena:2016hyu, Berkooz:2016cvq, Jian:2017unn,Gu_2017} and CFTs \cite{Roberts:2014ifa,Jackson_2015}, equation \eqref{otoceq} for $N$ number of d.o.f. per unit volume becomes,
\begin{equation}\label{eq:otocex}
    C(t,x) =\frac{1}{N}e^{\lambda_L\left( t-\frac{|x|}{v_{B}}\right)},
\end{equation} 
%where $v_B$ is known as the `butterfly velocity' describing the growth of the operator $W$ in the physical space. It also ascertains that there is an additional delay in scrambling due to the physical seperation between the two operators. 
where, $\lambda_{L}$ is the Lyapunov exponent that determines the overall growth of the correlator and  for maximally chaotic systems, it saturates the bound $\lambda_{L} \leq 2 \pi/\beta$ in natural units \cite{Maldacena_2016}. More specifically, for a large class of holographic theories with black holes in the AdS bulk, this bound saturates \cite{Grozdanov:2017ajz,Blake:2018leo,Grozdanov:2018kkt,Mahish:2022xjz}. Also, $v_B$ is known as the `butterfly velocity' which characterizes the growth of any early perturbation $W$ as discussed above. Moreover, there exists an effective light cone structure in spacetime as defined by the butterfly velocity such that any latter measurement in the form of an operator $V$, if inserted outside the light cone, will be unaffected by the early perturbation $W$, resulting in $C(t,x)=0$ \cite{Mezei:2019dfv,Mezei:2016wfz,Xu:2022vko,Roberts:2016wdl,Chakrabortty:2022kvq,Sil:2020jhr}.

The late time behavior of the $\mathrm{OTOC}$s, as given in equation (\ref{eq:otocex}), is in general appropriate when the corresponding local operators in the correlation function are not conserved. Instead, for operators which follow a conservation law, the $\mathrm{OTOC}$ at late time behaves quite differently.  
%In this paper we are interested in the late time (i.e. time much larger than the scrambling time) behaviour of $\mathrm{OTOC}$s. In general, following \eqref{eq:otocex}, the $\mathrm{OTOC}$ between two non--conserved operators decays to zero exponentially fast.
In recent years, it has been observed that for random circuit model \cite{Khemani:2017nda,Nahum_2018,Rakovszky_2018} and also for certain holographic theories \cite{Cheng:2021mop}, the $\mathrm{OTOC}$ between the conserved charge density operator and a non--conserved operator displays a diffusive power law tail at late times. This diffusion arises because the conserved charge density spreads over time, leading to a slower relaxation process. In their paper \cite{Cheng:2021mop}, Cheng \& Swingle discussed the power law fall--off for boundary $\mathrm{OTOC}$ involving $U(1)$ conserved charges which in the bulk corresponds to Maxwell gauge field $A_\mu$ in $(d+2)$--dimensional AdS--Schwarzschild black hole background. In this paper, we extend their work by considering higher--form fields in the bulk which by the holographic principle is dual to conserved current operators associated with higher--form global symmetries of the boundary theory.

Higher--form symmetries are symmetries whose charged operators have support on extended objects such as lines, surfaces, or other higher dimensional geometries (for a pedagogical review see \cite{Gomes:2023ahz}). Ordinary symmetries, the ones often discussed in QFT, are particular examples of higher--form symmetries (also called 0--form symmetries), as their charged operators are localized on zero--dimensional objects (point particles). The non--local nature of the charged objects poses a significant challenge in performing computations in these field theories. However, if these field theories have holographic duals then we can perform computations in the bulk with local bulk fields and relate it to say the correlators in the boundary. In this paper, we are primarily focused on antisymmetric $2$--form fields, $B_{\mu \nu}$ in the bulk with the usual 1--form $U(1)$ gauge symmetry which corresponds to a 1--form global symmetry at the boundary \cite{Hofman:2017vwr, Gaiotto:2014kfa, DeWolfe:2020uzb} \footnote{The $B_{\mu \nu}$ field can have 2--form global currents corresponding to conservation of field lines akin to 1--form electric and magnetic currents of the Maxwell gauge field $A_\mu$ \cite{Hofman:2017vwr}.}. Later, we have also discussed possible generalizations to $p$--form fields 
%in the bulk of arbitrary dimensions 
associated with $(p-1)$--form global $U(1)$ symmetries in the boundary theory. We have computed the boundary  $\mathrm{OTOC}$s involving the higher--form currents in the scattering approach \cite{Shenker:2014cwa}, as inner product between two asymptotic `in' and `out' states. Since these states are related to the bulk fields via the bulk--to--boundary propagators  \cite{Shenker:2014cwa,Cheng:2021mop}, $\mathrm{OTOC}$s are equivalent to the inner product between two bulk fields with and without a shockwave. We have observed that at late times $\mathrm{OTOC}$s always display power law tails irrespective of the nature of the form fields; it is a universal feature. Another important linear response phenomena related to the non--uniqueness of correlation functions at some special points, is known as `pole--skipping'. In \cite{Wang:2022xoc}, pole--skipping properties of higher--forms fields with $U(1)$ gauge symmetries in the bulk is discussed in detail and thus will not be addressed further in this paper.

The rest of the paper is organized as follows. We briefly review the holographic computation of the $\mathrm{OTOC}$ in section \ref{otocreview}. In section \ref{sec:2formfields}, we discuss the case of a $B$--field, propagating in AdS--Schwarzschild geometry. In five spacetime dimensions $B$--field exhibits logarithmic divergence and needs regularization. In section \ref{sec:2formotoc5dim}, the $\mathrm{OTOC}$ between a conserved $U(1)$ charge operator for the $B$--field and a heavy scalar operator is discussed. The late time power law behavior of the $\mathrm{OTOC}$ is confirmed by equating it to the inner product of $B$--fields in a shockwave geometry resulting from high energy heavy scalar operator quanta. In section \ref{sec:otherdimotherform}, we generalize our results from the previous sections to higher--form fields in arbitrary spacetime dimensions. In appendix \ref{sec:b7dim}, we discuss $B$--fields in six and seven dimensions where no boundary divergence is observed. 
%In appendix \ref{abduality}, we show some equivalences in terms of the boundary regularization and the diffusion constants between $A_\mu$ and $B$--fields in different spacetime dimensions. 
Finally, we summarize with a brief conclusion.

\section{$\mathrm{OTOC}$ in holography: A brief review} \label{otocreview}
The study of Out--of--Time-Ordered Correlators ($\mathrm{OTOC}$s) within the framework of holography has become a key approach to understanding quantum chaos and information scrambling. Over the last decade or so,  research in the directions of quantum gravity, field theory and quantum many body systems has thoroughly explored the holographic computation of OTOCs, particularly for scalar operators \cite{Shenker:2014cwa,Shenker:2013pqa,Roberts:2014ifa,Shenker:2013yza,Roberts:2014isa,Cheng:2021mop}. In this section, we adopt a scattering based approach \cite{Shenker:2014cwa,Cheng:2021mop}, where the boundary OTOCs are expressed as inner products between `in' and `out' asymptotic states, offering a valuable method for probing the chaotic dynamics in strongly coupled systems.
 The $\mathrm{OTOC}$s have different versions but a particularly interesting one is the Left ($L$), Right ($R$) version represented as,
\begin{equation}
\langle V_L(t_2, \vec{x}_2) W_R(t_1, \vec{x}_1) V_R(t_2, \vec{x}_2)  W_L(t_1, \vec{x}_1)   \rangle = \langle \mathrm{out} | \mathrm{in} \rangle \, ,
\end{equation}
where the states \( |\mathrm{in}\rangle \) and \( |\mathrm{out}\rangle \) correspond to the time evolved combinations of the operators \( V \) and \( W \), acting on a thermofield double (TFD) state. Specifically, these states are given by,
\begin{equation}
\begin{aligned}
&
|\mathrm{in} \rangle =  V_R(t_2, \vec{x}_2) W_L(t_1, \vec{x}_1)|\text{TFD} \rangle
\\&
|\mathrm{out} \rangle =  W_R^\dagger(t_1, \vec{x}_1) V_L^\dagger(t_2, \vec{x}_2)|\text{TFD} \rangle \, .
\end{aligned}
\end{equation}
The \( |\text{TFD} \rangle \) state, plays a central role in this setup. It is created by entangling two identical copies of a conformal field theory (CFT) and can be written as,
\begin{equation}
|\text{TFD} \rangle = \frac{1}{\sqrt{Z(\beta)}} \sum_n e^{-\frac{\beta E_n}{2}} |E_n \rangle_L |E_n \rangle_R \, .
\end{equation}
Here, \( |E_n \rangle_L \) and \( |E_n \rangle_R \) are the energy eigenstates of the left and right CFTs, respectively and \( Z(\beta) \) is the partition function at inverse temperature \( \beta \). In holography, this thermofield double state corresponds to an eternal black hole in anti--de Sitter (AdS) space, providing a geometric dual for the quantum state.
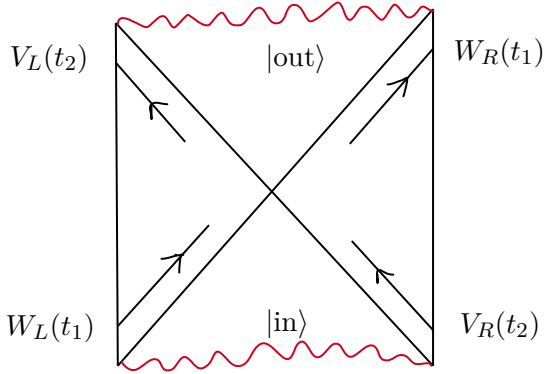
\begin{figure}[h!]
		\centering

    \tikzset{every picture/.style={line width=0.75pt}} %set default line width to 0.75pt        

\begin{tikzpicture}[x=0.75pt,y=0.75pt,yscale=-1,xscale=1]
\path(0,200);

%Straight Lines [id:da7229253864197858] 
\draw    (200,37) -- (201,210) ;
%Straight Lines [id:da9347370150642308] 
\draw    (360,30) -- (360,210) ;
%Straight Lines [id:da18762923001273868] 
\draw    (360,30) -- (201,210) ;
%Straight Lines [id:da4427847417709402] 
\draw    (360,210) -- (200,37) ;
%Shape: Free Drawing [id:dp48449251331879983] 
\draw  [color={rgb, 255:red, 208; green, 2; blue, 27 }  ,draw opacity=1 ][line width=0.75] [line join = round][line cap = round] (201,38) .. controls (201,29) and (213.59,40.38) .. (216,39) .. controls (218.98,37.3) and (222.36,31.8) .. (225,34) .. controls (226.81,35.51) and (228.21,40.53) .. (230,39) .. controls (232.15,37.16) and (233.85,34.84) .. (236,33) .. controls (237.16,32) and (243.23,39.41) .. (247,38) .. controls (248.03,37.61) and (252.42,30.42) .. (254,32) .. controls (256.85,34.85) and (256.36,39.11) .. (263,38) .. controls (266.4,37.43) and (269.83,26.42) .. (274,30) .. controls (276.56,32.2) and (277.27,38.18) .. (282,37) .. controls (286.27,35.93) and (285.46,28.88) .. (289,28) .. controls (295,26.5) and (295.13,37.55) .. (299,36) .. controls (303.95,34.02) and (306.76,29.82) .. (311,27) .. controls (315.01,24.33) and (315.08,37.94) .. (319,35) .. controls (321.76,32.93) and (322.12,27.37) .. (327,29) .. controls (329.31,29.77) and (332.97,37.03) .. (335,35) .. controls (337.03,32.97) and (337.97,30.03) .. (340,28) .. controls (341.79,26.21) and (344.8,33.12) .. (347,34) .. controls (350.19,35.28) and (352.66,29.96) .. (354,29) .. controls (355.79,27.72) and (357.87,31) .. (360,31) ;
%Shape: Free Drawing [id:dp05011670582074512] 
\draw  [color={rgb, 255:red, 208; green, 2; blue, 27 }  ,draw opacity=1 ][line width=0.75] [line join = round][line cap = round] (203,208) .. controls (202.64,208) and (206.09,213.17) .. (210,212) .. controls (213.86,210.84) and (217.4,205.2) .. (221,207) .. controls (223.53,208.26) and (222.59,213.8) .. (228,212) .. controls (230.85,211.05) and (232.48,203.65) .. (236,206) .. controls (239.81,208.54) and (241.26,214.74) .. (247,209) .. controls (248.84,207.16) and (249.67,201.84) .. (252,203) .. controls (255.7,204.85) and (257.87,211.28) .. (263,210) .. controls (268.11,208.72) and (268.25,201.38) .. (273,199) .. controls (278.41,196.29) and (284.32,209.68) .. (288,206) .. controls (290.22,203.78) and (290.19,199.41) .. (293,198) .. controls (295.95,196.52) and (297.52,202.83) .. (300,205) .. controls (303.7,208.24) and (307.24,199.69) .. (310,199) .. controls (314.08,197.98) and (313.63,202.47) .. (317,205) .. controls (320.6,207.7) and (325.74,197.87) .. (330,200) .. controls (333.02,201.51) and (332.03,206.01) .. (335,207) .. controls (339.9,208.63) and (341.27,202.9) .. (346,205) .. controls (347.95,205.87) and (348.96,208.37) .. (351,209) .. controls (353.57,209.79) and (359,207.31) .. (359,210) ;
%Straight Lines [id:da7348865329411891] 
\draw    (247,139) -- (201,190) ;
%Straight Lines [id:da058301336022897] 
\draw    (200,57) -- (235,97) ;
%Straight Lines [id:da06455753400033748] 
\draw    (360,50) -- (318,98) ;
%Straight Lines [id:da5575885976702107] 
\draw    (360,192) -- (319,147) ;
%Shape: Free Drawing [id:dp8924684114414316] 
\draw  [color={rgb, 255:red, 0; green, 0; blue, 0 }  ,draw opacity=1 ][line width=0.75] [line join = round][line cap = round] (332,164) .. controls (332,165.33) and (332,169.33) .. (332,168) .. controls (332,166) and (331.37,163.9) .. (332,162) .. controls (332.21,161.37) and (333.4,161.7) .. (334,162) .. controls (335.61,162.8) and (336.88,164) .. (340,164) ;
%Shape: Free Drawing [id:dp41784574210540326] 
\draw  [color={rgb, 255:red, 0; green, 0; blue, 0 }  ,draw opacity=1 ][line width=0.75] [line join = round][line cap = round] (337,68) .. controls (339.49,68) and (341.82,66.55) .. (344,66) .. controls (345.02,65.74) and (347,63.95) .. (347,65) .. controls (347,67.56) and (345,74.5) .. (345,74) ;
%Shape: Free Drawing [id:dp34955119910669696] 
\draw  [color={rgb, 255:red, 0; green, 0; blue, 0 }  ,draw opacity=1 ][line width=0.75] [line join = round][line cap = round] (218,79) .. controls (218,81) and (218,87) .. (218,85) .. controls (218,82.67) and (216.24,79.54) .. (218,78) .. controls (220.01,76.24) and (223.33,78) .. (226,78) ;
%Shape: Free Drawing [id:dp19864934648855326] 
\draw  [color={rgb, 255:red, 0; green, 0; blue, 0 }  ,draw opacity=1 ][line width=0.75] [line join = round][line cap = round] (537,150) .. controls (537,151) and (537,152) .. (537,153) ;
%Shape: Free Drawing [id:dp4875677785608763] 
\draw  [color={rgb, 255:red, 0; green, 0; blue, 0 }  ,draw opacity=1 ][line width=0.75] [line join = round][line cap = round] (230,154) .. controls (230,154) and (230,154) .. (230,154) ;
%Shape: Free Drawing [id:dp45470016726938545] 
\draw  [color={rgb, 255:red, 0; green, 0; blue, 0 }  ,draw opacity=1 ][line width=0.75] [line join = round][line cap = round] (223,156) .. controls (227,156) and (229.41,154) .. (234,154) .. controls (234.33,154) and (233.11,153.68) .. (233,154) .. controls (232.19,156.43) and (232,160.49) .. (232,163) ;

% Text Node
\draw (369,43) node [anchor=north west][inner sep=0.75pt]   [align=left] {$W_R(t_1)$};
% Text Node
\draw (274,46) node [anchor=north west][inner sep=0.75pt]   [align=left] {$|\mathrm{out}\rangle$};
% Text Node
\draw (372,181) node [anchor=north west][inner sep=0.75pt]   [align=left] {$V_R(t_2)$};
% Text Node
\draw (145,46) node [anchor=north west][inner sep=0.75pt]   [align=left] {$V_L(t_2)$};
% Text Node
\draw (143,182) node [anchor=north west][inner sep=0.75pt]   [align=left] {$W_L(t_1)$};
% Text Node
\draw (274,180) node [anchor=north west][inner sep=0.75pt]   [align=left] {$|\mathrm{in}\rangle$};
\end{tikzpicture}
\caption{The `in' and `out' states in the Penrose diagram}
\end{figure}

In the bulk picture, the `in' and `out' states are related to the bulk wavefunctions \( \phi_V \) and \( \phi_W \), which correspond to the boundary CFT operators \( V \) and \( W \) ,
\begin{equation}
\begin{aligned}
| \mathrm{in}\rangle & =\int d p^v d \vec{x} \int d p^u d \vec{x}^{\prime} \psi_W\left(p^v, \vec{x}\right) \psi_V\left(p^u, \vec{x}^{\prime}\right)\left|p^v, \vec{x}; p^u, \vec{x}^{\prime}\right\rangle_{\mathrm{in }} \\
|\mathrm{out}\rangle & =\int d p^v d \vec{x} \int d p^u d \vec{x}^{\prime} \psi_{W^{\dagger}}\left(p^v, \vec{x}\right) \psi_{V^{\dagger}}\left(p^u, \vec{x}^{\prime}\right)\left|p^v, \vec{x}; p^u, \vec{x}^{\prime}\right\rangle_{\mathrm{out }} \,.
\end{aligned}
\end{equation}
The coordinates, \( u \) and \( v \) are the null Kruskal coordinates of the black hole geometry dual to the $|\text{TFD} \rangle$. The scattering of these wavefunctions near the approximately flat near horizon region of the black hole becomes the main mechanism for calculating the OTOC. The energy scale of the scattering process is determined by the Mandelstam variable \( s = 2 p_1^v p_2^u \sim e^{2\pi t_{12} / \beta} \), where \( t_{12} \) is the large time separation between the two boundary operators. Using the standard AdS/CFT dictionary \cite{Maldacena:1997re,maldacena2014gaugegravityduality,McGreevy_2010,ramallo2013introductionadscftcorrespondence,horowitz2006gaugegravityduality,Hubeny_2015,witten1998antisitterspaceholography}, the bulk wavefunctions near the boundary sourced by the boundary operators can be written in terms of bulk--to--boundary propagators as, %(`extrapolate' dictionary),
\begin{equation}
\begin{aligned}
\psi_W\left(p_1^v, \vec{x}\right) & =\left.\int d u ~e^{i p_1^v u}\left\langle\psi_W(u, v, \vec{x}) W_L\left(t_1, \vec{x}_1\right)\right\rangle\right|_{v=0} \\
\psi_V\left(p_2^u, \vec{x}^{\prime}\right) & =\left.\int d v ~e^{i p_2^u v}\left\langle\psi_V(u, v, \vec{x}^{\prime}) V_R\left(t_2, \vec{x}_2\right)\right\rangle\right|_{u=0} \\
\psi_{W^{\dagger}}\left(p_1^v, \vec{x}\right) & =\left.\int d u ~e^{i p_1^v u}\left\langle\psi_W(u, v, \vec{x}) W_R^{\dagger}\left(t_1, \vec{x}_1\right)\right\rangle\right|_{v=0} \\
\psi_{V^{\dagger}}\left(p_2^u, \vec{x}^{\prime}\right) & =\left.\int d v ~e^{i p_2^u v}\left\langle\psi_V(u, v, \vec{x}^{\prime}) V_L^{\dagger}\left(t_2, \vec{x}_2\right)\right\rangle\right|_{u=0}\,.
\end{aligned}
\end{equation}

For scrambling physics, we are interested in time scales that are larger than the relaxation time and the dynamics is dominated by scattering processes near the black hole horizon. To the leading order in $s$ and under the eikonal gravity approximation \cite{Shenker:2014cwa}, the S--matrix approaches a pure phase,
\begin{equation}
| p^v_1, \vec{x}, p^u_2, \vec{x}' \rangle_{\text{out}} \sim e^{i\delta(s, b)} | p^v_1, \vec{x}, p^u_2, \vec{x}' \rangle_{\text{in}} + |\chi \rangle
\end{equation}
where \( b = |\vec{x} - \vec{x}'| \) is the transverse separation between the particles, and the state $ |\chi\rangle$ represents the inelastic component of the scattering and is orthogonal to all `in' states that consist of a single $W$  particle and a single $V$ particle. The phase shift \( \delta(s, b) \) accumulated due to the scattering arises from the interaction of particle/quanta of operator $V_R$ with the gravitational shockwave (located near $u\sim 0$) sourced by particle/quanta of the operator $W_L$ as it crosses the black hole horizon \cite{Dray:1985yt,DRAY1985173,Shenker:2014cwa}. 

The shockwave introduces a displacement in the \( v \) coordinate of the bulk wavefunction such that the OTOC is modified and in the case for large time separation, it decays exponentially. Specifically, in systems where the operator \( W \) is much heavier than \( V \) (i.e. \( \Delta_W \gg \Delta_V \gg 1 \)), the wavefunction of the heavy $W$ particles remain largely unaffected and the OTOC simply reduces to an inner product between the lighter $V$ particle wavefunctions before and after the shockwave, 
\begin{equation}
\text{OTOC} \sim \int dv d\vec{x} \psi^L_V(v, \vec{x}) \partial_v \psi^R_V(v - h(\vec{x}), \vec{x})\, .
\end{equation}
Here, $\displaystyle h(\vec{x}) = G_N p_1^v \frac{e^{-\mu |\vec{x}|}}{|\vec{x}|^{(d-1)/2}}$ is the shift in the $v$ coordinate for $u>0$, $ \displaystyle \mu = \sqrt{\frac{2d}{d+1}} \frac{2\pi}{\beta}= \frac{2\pi}{v_B \beta}$ controls the exponential falloff and $v_B$ is the `butterfly velocity'. This yields the late time behavior of the OTOC as
\begin{equation}\label{eq:otocexp}
\text{OTOC} \sim \left[ \frac{1}{1 + \frac{G_N \Delta_W}{\Gamma} e^{2\pi t / \beta - \mu |\vec{x}|}} \right]^{\Delta_V}\, .
\end{equation}
The parameter \( \Gamma \) depends on the regularization of the correlator \cite{Shenker:2014cwa}. %\cite{Grozdanov:2017ajz}.%
In chaotic systems, this formula results in an exponential decay of the OTOC at late times. 

\subsection*{$\mathrm{OTOC}$ and conservation law}

In systems with a conserved charge, the OTOC decays more slowly, exhibiting a diffusive power--law tail \cite{Khemani:2017nda,Nahum_2018,Rakovszky_2018}. Due to the hydrodynamical property of the conserved current, the particle sourced by these operators in the bulk spreads over a large region of spacetime. Consequently, in the classical picture, the collision that causes scrambling is spread over a large range of spacetime points. The center of mass energy is large when the collision happens close to the horizon and gets smaller farther away from the horizon. This leads to a smearing of the exponential
factor in \eqref{eq:otocexp}, effectively replacing the OTOC formula with
\begin{equation}
\text{OTOC} \sim \int_0^\infty ds \frac{1}{s^{d/2+1}} \left[ \frac{1}{1 + \frac{c}{N} e^{2\pi (t - s) / \beta - \mu |\vec{x}|}} \right]^\alpha,
\end{equation}
where $\alpha$ and $c$ are constants. For gauge fields, this expression leads to a power law decay at late times, proportional to \( t^{-d/2} \) \cite{Cheng:2021mop}. The slower decay is a hallmark of systems governed by conservation laws, where the dynamics are dominated by diffusive processes. In this work, we extend these observations to higher--form symmetries, showing that similar power law behavior emerges universally in systems with U(1) charge conservation.

\section{$B_{\mu\nu}$ field in AdS--Schwarzschild geometry} \label{sec:2formfields}

It is possible to study higher--form symmetries in terms of the dual gravitational AdS bulk through the lenses of the holographic principle. Such higher--form global symmetries at the boundary are associated with the antisymmetric tensor gauge fields in the AdS bulk and arise from the existence of completely antisymmetric conserved currents $J^{[\mu \nu \dots]}$ \cite{Gaiotto:2014kfa,Hofman:2017vwr,DeWolfe:2020uzb,Gomes_2023}. The currents being differential forms allow for the construction of topological conserved charges by integrating over specific lower--dimensional spatial manifolds and as a result the charged objects are also non--local operators extended over lines, surfaces etc. Higher--form symmetries are prevalent in various relativistic theories, including both abelian and non--abelian gauge theories and belong to an even wider class of symmetries referred to as generalized symmetries \cite{Sharpe_2015,Bhardwaj:2023kri,Brennan:2023mmt}. The intensive study of these symmetries has led to a powerful, integrated framework, drawing ideas from quantum field theory, topological phases of matter, string theory, quantum computing, and even quantum gravity. An illustrative example of generalized symmetries can be seen in gauge theories, where charged objects such as Wilson and 't Hooft operators serve as essential tools for probing the topological and geometric properties of these theories.

%\st{Higher--form symmetries are prevalent in various relativistic theories, including both abelian and non--abelian gauge theories. These symmetries arise from the existence of completely antisymmetric conserved currents $J^{[\mu \nu \dots]}$ \cite{Gomes_2023}. The currents being differential forms allow for the construction of conserved charges by integrating over specific lower--dimensional spatial manifolds, rather than the entire space as is typical for ordinary (0--form) symmetries, where particles serve as the natural charged entities. In contrast, for higher--form symmetries, the charged objects are not local operators; they are extended entities e.g., lines, surfaces, etc. These objects are challenging to handle, and a structured approach to developing these theories is yet to be established. However, through the lenses of the holographic principle it is possible to study these higher--form symmetries in terms of the dual gravitational AdS bulk. Such higher--form global symmetries in the boundary is associated with the antisymmetric tensor gauge fields in the AdS bulk \cite{Gaiotto:2014kfa,Hofman:2017vwr,DeWolfe:2020uzb}.} 

Our focus is primarily on the bulk side, where antisymmetric tensor gauge fields are form fields realized through differential forms and appear naturally in the string spectrum. The current section is devoted to the study of 2--form, $B$--fields in the AdS--Schwarzschild geometry in various spacetime dimensions which is later expanded to higher--form fields in section \ref{sec:otherdimotherform}. The analysis for the 1--form Maxwell field $A_\mu$ is well studied in AdS/CFT \cite{Cheng:2021mop,Faulkner:2012gt,Wang:2022xoc}. %A brief discussion on Maxwell field $A_\mu$ can be found in appendix A. 
The differences between different form fields are mainly sourced from their boundary conditions. As we shall see, this results in different boundary interpretations. In section \ref{sec:2formotoc5dim}, we shall make use the results from the current section to arrive at the late time behavior of $\mathrm{OTOC}$s by studying the effect of shockwave geometry on these fields.

\subsection{AdS--Schwarzschild black hole}
We consider AdS--Schwarzschild black hole spacetime in $d+2$ dimensions with $d$ number of transverse directions. In $d+2$ dimension, the metric of the black hole can be written as,
\begin{equation}
d s^2=\frac{L^2}{z^2} \left[-f(z) d t^2+\frac{1}{f(z)} d z^2+ (dx_1^2 +\dots+dx_d^2)\right] \, ,\quad f(z)=1-\left(\frac{z}{z_{H}}\right)^{d+1}\, .
\end{equation}
In five spacetime dimensions it takes the form,
\begin{equation}
d s^2=\frac{L^2}{z^2} \left[-f(z) d t^2+\frac{1}{f(z)} d z^2+ (dx^2 +dy^2 +dw^2 )\right] \, ,\quad f(z)=1-\left(\frac{z}{z_{H}}\right)^{4}\, ,
\end{equation}
where $z=z_{H}$ is the horizon and the boundary is at $z=0$. The inverse temperature associated with the horizon is $\displaystyle \beta=\pi z_{H}$. The transverse directions are labeled by the coordinates $x$, $y$ and $w$. To perform the $\mathrm{OTOC}$ computations, it is convenient to switch to the tortoise coordinates. Hence we define $r_*$ according to, 
\begin{equation}
d r_*=-\frac{1}{f(z)} d z \,  ,
\end{equation}
where $r_* \in(-\infty, 0]$ with $-\infty$ corresponding to the horizon and the boundary is at $r_*=0$. In these coordinates, the relevant metric components are, $\displaystyle g_{r_* r_*}= -g_{t t}=L^2 \frac{f(z)}{z^2}$ and the metric takes a more symmetric form, 
\begin{equation}
d s^2=\frac{L^2}{z^2} \left[-f(z) d t^2+f(z) d r_*^2+ (dx^2 +dy^2 +dw^2 )\right].
\end{equation}
For the rest of the paper, we set the AdS radius, $L=1$ . 
\subsection{$B$--field in five dimensional AdS--Schwarzschild geometry}\label{sec:b5dim}
The 2--form field $B_{\mu \nu}$ is totally antisymmetric in its two indices and the corresponding field strength $H=dB$ is a 3--form.
A minimally coupled 2--form field $B$ has the action \cite{Hofman:2017vwr,Gomes_2023},
\begin{equation} \label{eq:bfieldaction}
S=\frac{1}{6 \gamma^2} \int \mathrm{d}^{5} x \sqrt{-g} H^2\,,
\end{equation}
where $\gamma$ is the coupling constant and the field strength $H$ has components $H_{\mu  \nu \rho}=\partial_\mu B_{\nu \rho}+\partial_\nu B_{\rho \mu}+\partial_\rho B_{\mu \nu}$ . The theory has a $U(1)$ gauge symmetry, $B_{\mu \nu} \rightarrow B_{\mu \nu}+\partial_\mu \Lambda_\nu-\partial_\nu \Lambda_\mu$, for any local vector $\Lambda$ . 
%This action \eqref{eq:bfieldaction} is Poincare dual of a Maxwell gauge field $A$, related to $B$ via $dB=\star_5 ~dA$ \cite{Hofman:2017vwr}.
Consider a $B$--field propagating in the five dimensional AdS--Schwarzschild geometry whose dynamics determines the behavior of a $U(1)$ current operator $J$ on the boundary. It can be derived from the on--shell action, 
\begin{equation}\label{eq:Jdef}
    J^{\mu \nu}=-\frac{1}{\gamma^2}\sqrt{-g} H^{r_{*} \mu \nu}\, ,
\end{equation}
and has a continuation in the bulk when necessary. In section \ref{sec:2formotoc5dim}, we shall calculate the four--point correlation function ($\mathrm{OTOC}$) with two insertions of $J$ and two insertions of a much heavier scalar operator $O$. In this section we shall focus on the the dynamics of the $B$--field in the blackhole geometry. 

\subsubsection{Equation of motion}
The equation of motion can be derived by taking the functional derivatives of the action \eqref{eq:bfieldaction},
\begin{equation} \label{eombmn}
\partial_\mu\left(\sqrt{-g} H^{\mu \nu \rho}\right)=\partial_\mu\left(\sqrt{-g} g^{\mu \alpha} g^{\nu \beta} g^{\rho \sigma} H_{\alpha \beta  \sigma}\right)= 0 \,. 
\end{equation}
%Here we will not fix any gauge. 
We are looking for solutions propagating along the $x$--direction \footnote{For each $\vec{q}$ , we pick an orthogonal coordinate frame such that the $x$--axis is parallel to $\vec{q}$ . While doing the computations we assumed the fields to be moving along the $x$--direction with momentum $q$. 
 %This choice is arbitrary; we can always rotate the momentum $\vec{q}$ along any direction by the rotation group. But, we shall keep using the vector notation for $\vec{x}$ and $\vec{q}$ as it only shows up through the Fourier term $e^{i\vec{q}\cdot\vec{x}}$. We can think of the direction $y$ being transverse to $\vec{q}$ .
 }. The field admits a plane--wave expansion,
\begin{equation}
B_{\mu \nu}(r_*, t, \vec{x})=\int d \vec{q} ~ d \omega  ~e^{-i \omega t+i \vec{q}\cdot \vec{x}} ~b_{\mu \nu} (r_*, \omega, \vec{q})\, .
\end{equation}
In the asymptotic AdS background, the spatial $SO(d-1)$ rotational symmetry yields decomposition of the field components into different modes, namely the scalar, the vector and the tensor modes
%\footnote{Scalar mode: $B_{tx}$, $B_{tr_*}$, $B_{xr_*}$ ; Vector mode: $B_{ty}$, $B_{r_* y}$, $B_{xy}$, $B_{tw}$ and $B_{xw}$ ; Tensor mode: $B_{yw}$ . The scalar mode computation is covered in Stanford's papers \cite{Shenker:2013pqa, Shenker:2014cwa} and their OTOCs decay exponentially at large time scales. The Tensor mode doesn't have any hydrodynamic structure at low frequencies and we shall not study it further \cite{Hofman:2017vwr} \AC{Is this correct?}.}
 \footnote{Because the current now has an extra spatial traverse index $y$, it is conceivable that the field only sees spatial $SO(d-2)$ rotational symmetry. In this case, the subsequent computations goes through as if $d\rightarrow (d-1)$ and is discussed in section \ref{sec:otherdimotherform}.}. We shall focus only on the vector channel as it has interesting hydrodynamic structure \cite{Hofman:2017vwr}. Moreover, we will consider the ansatz such that all $B_{\mu \nu}$ field components except $B_{ty}$ and $B_{r_*y}$ are zero \footnote{If some of the field we are setting to zero are in fact not zero, it would reflect as inconsistencies while solving the equations of motion.}. With this ansatz, we have the following two independent equations,
\begin{equation} \label{eq:beom1}
 b_{r_*y}(r_*) \left(q^2 f(r_*)-\omega^2\right)+i \omega b_{ty}'(r_*) = 0
\end{equation}
\begin{equation} \label{eq:beom2}
\omega b_{ty}(r_*) z(r_*)-i \left(z(r_*) b_{r_*y}'(r_*)+b_{r_*y}(r_*) z'(r_*)\right)= 0 \,,
\end{equation}
 where prime (${}^\prime$) denotes derivative with respect to $r_*$. To solve these coupled differential equations, it is convenient to substitute $b_{r_*y}$ from \eqref{eq:beom1} in equation \eqref{eq:beom2} to get a second order differential equation in $b_{ty}$,

\begin{comment}
\begin{equation}
    b_{ty}''(r_*)+  \left(\frac{q^2 f'(r_*)}{\omega^2-q^2 f(r_*)} +\frac{z'(r_*)}{z(r_*)}\right) b_{ty}'(r_*) +  \left(\omega^2-q^2 f(r_*)\right) b_{ty}(r_*) = 0
\end{equation}
\end{comment}

\begin{equation} \label{eq:emain}
    b_{ty}''(r_*)- \partial_{r_*} \text{ln} \left(\frac{\omega^2-q^2 f(r_*)}{z(r_*)}\right)~ b_{ty}'(r_*) +  \left(\omega^2-q^2 f(r_*)\right) b_{ty}(r_*) = 0 \, .
\end{equation}
At the horizon $z=z_H$, $f(r_*)=0$ and above equation reduces to 
\begin{equation}
    b_{ty}''(r_*)+ \omega^2 ~b_{ty}(r_*) = 0.
\end{equation}
Considering only the in-falling mode, the ansatz near the horizon is given as,   
\begin{equation}
b_{ty}(r_*, \omega, q) = e^{-i \omega r_*} F(r_*,\omega,q) \,.
\end{equation}
It might be possible to solve for $F(r_*)$ in more general terms, but we are primarily interested in the hydrodynamic limit i.e. we scale $\omega$ and $q$ by $\lambda$ with $\lambda \ll 1$ and consider the following perturbative expansion of $F$,
\begin{equation}
 F=F_0+\lambda F_1+ \lambda^2 F_2 \,+  \cdots\,,
\end{equation}
ultimately solving the second order differential equation \eqref{eq:emain} up to first order in $\lambda$. We note that the scaling by $\lambda$ is a convenient bookkeeping tool to keep track of the powers of $\omega$, $q$ and the combinations thereof. 
%\subsubsection{Solutions to the E.O.M.}
%We shall now solve the second order differential equation \eqref{eq:emain} order by order in $\lambda$.

\begin{comment}
\begin{itemize}
    \item \textbf{The Scalar Mode}: $B_{tx}$ .
    \item \textbf{The Vector Mode} (Longitudinal Mode): $B_{tr}$, $B_{xr}$, $B_{ty}$, $B_{xy}$, $B_{tw}$ and $B_{xw}$ \,.
    \item \textbf{The Tensor Mode} (Transverse Mode): $B_{yw}$. It does not have any hydrodynamic structure at low frequencies and we will not study it further \cite{Hofman:2017vwr}. 
\end{itemize}
\end{comment}

\subsubsection*{\textbf{Zeroth order $\lambda^0$:}}
The differential equation takes the form,
\begin{equation}
    F_{0}''(r_*)+  \left(\frac{q^2 f'(r_*)}{\omega^2-q^2 f(r_*)}+\frac{z'(r_*)}{z(r_*)}\right) F_{0}'(r_*) = 0 \, ,
\end{equation}
and the solution is

\begin{comment}
\begin{equation}
    F_{0}'(r_*)= C_0 \frac{ \left(\omega^2-q^2 f(r_*)\right)}{z(r_*)}
\end{equation}
\end{comment}

\begin{equation}
    F_{0}(r_*)= \int d r_* ~C_0 \frac{ \left(\omega^2-q^2 f(r_*)\right)}{z(r_*)} + C \, .
\end{equation}
Integration is straightforward and it gives 
\begin{equation}
F_{0}(z)= C_0 \left(\left(q^2-\omega^2\right) \ln (z)+\frac{1}{4} \omega^2 \,\ln\left(z^4-z_{H}^4\right)\right)+ C\, .
\end{equation}
The regularity at the horizon $z=z_H$ sets $C_0$ to zero. Therefore, to zeroth order in $\lambda$ we have $F_0(r_*)=C$.

\subsubsection*{\textbf{First order $\lambda^1$:}}
The differential equation \eqref{eq:emain} to the first order in $\lambda$ after making the substitution for the zeroth order $F_0(r_*)=C$, is given as,

\begin{comment}
\begin{equation}
    F_{1}''(r_*)-2 i \omega F_{0}'(r_*)+\left(F_{1}'(r_*)-i \omega F_{0}(r_*)\right) \left(\frac{q^2 f'(r_*)}{\omega^2-q^2 f(r_*)}+\frac{z'(r_*)}{z(r_*)}\right)=0 \, ,
\end{equation}
\end{comment}

\begin{equation}
F_{1}''(r_*)+\left(F_{1}'(r_*)-i \omega C \right) \left(\frac{q^2 f'(r_*)}{\omega^2-q^2 f(r_*)}+\frac{z'(r_*)}{z(r_*)}\right)=0 \, .
\end{equation}
Again, imposing the regularity at the horizon, $F_{1}'(z=z_H)=0$, we obtain the solution for $F_1$ as,

\begin{comment}
\begin{equation}
    F_{1}'(r_*)=i \omega C \left[1 - \left(1-\frac{q^2 f(r_*)}{\omega^2}\right) \frac{z_H}{z(r_*)}\right]
\end{equation}
\end{comment}

\begin{equation}
    F_{1}(r_*)=i \omega C \int_{-\infty}^{r_*} d r_* \left[1 - \left(1-\frac{q^2 f(r_*)}{\omega^2}\right) \frac{z_H}{z(r_*)}\right] \,.
\end{equation}

\subsubsection*{\textbf{The full solution upto $\mathcal{O}(\lambda)$}:}
Substituting the results for $F_0$ and $F_1$ the final solution for $F(r_*)$ up to first order in $\lambda$, takes the following form, 
\begin{equation}
    F=C \left[1+  i \omega  \int_{-\infty}^{r_*} d r_* \left(1 - \left(1-\frac{q^2 f(r_*)}{\omega^2}\right) \frac{z_H}{z(r_*)}\right) +\mathcal{O}\left(\lambda^2\right) \right]\, ,
\end{equation}
and subsequently $b_{ty}$ takes the form,
%\begin{equation}
 %   b_{ty}=e^{-i \omega r_*} ~C(\omega,q) \left[1+  i \omega  \int_{-\infty}^{r_*} d r_* \left(1 - \left(1-\frac{q^2 f(r_*)}{\omega^2}\right) \frac{z_H}{z(r_*)}\right)  \right]
%\end{equation}
%or,
\begin{equation}\label{eq:btywithc}
\begin{aligned}
   b_{ty}=e^{-i \omega r_*} ~C(\omega,q) & \left[ 1+  i \omega  \int_{-\infty}^{r_*} d r_* \left(1 - \frac{z_H}{z(r_*)}\right) \right.\\
   & \quad \left.+ \frac{i q^2}{\omega} \int_{-\infty}^{r_*} d r_*  f(r_*) \frac{z_H}{z(r_*)}  +\mathcal{O}\left(\lambda^2\right) \right] \, .
\end{aligned}
\end{equation}
We can fix the normalization constant by demanding  $\lim_{r_*\to 0} b_{ty}=1$. Then the normalized expression becomes,
\begin{equation} \label{eq:btyfinal1withCwk}
   b_{ty} (r_*, \omega, q)=e^{-i \omega r_*} \left[ \frac{1+ i \omega H(r_*)+\frac{i q^2}{\omega} \int_{-\infty}^{r_*} d r_*  f(r_*) \frac{z_H}{z(r_*)} }{1+ i \omega H(0)+\frac{i q^2}{\omega} \int_{-\infty}^{0} d r_*  f(r_*) \frac{z_H}{z(r_*)} }+\mathcal{O}\left(\lambda^2\right)\right] \,,
\end{equation}
where $H(r_*)$ is defined as,
\begin{equation}
   H(r_*)= \int_{-\infty}^{r_*} d r_* \left(1 - \frac{z_H}{z(r_*)}\right) \, .
\end{equation}

The indefinite integral $H(r_*)$ can be neglected. As we shall see in section \ref{sec:otherdimotherform}, the existence of this factor depends on the type of form fields and the dimensionality of spacetime in which they are  propagating. However, if we ignore the $\omega H(0)$ term in the denominator, then \eqref{eq:btyfinal1withCwk} has a pole at $\displaystyle \omega=-i  q^2 \int_{-\infty}^{0} d r_*  f(r_*) \frac{z_H}{z(r_*)}=-i  q^2 D$, where $D$ is the diffusion constant, see section \ref{sec:diff} . The dispersion relation now scales as $\omega \sim q^2$ and  $\omega H(0)$ is indeed subleading compared to the $i q^2 D / \omega$ term at small $\omega$ and $q$. 
%We can anticipate that $\omega^2$ is of order $q^4$ after using the residue theorem. Hence, to the leading order in small $q$ and $\omega$, we can neglect $\omega^2$ and higher order terms. 
Another useful approximation is to set $\displaystyle \left(\frac{z}{z_{H}}\right) \sim 1$, corresponding to the near horizon region. Since the scrambling time is large, the relevant physics indeed happens within this region. With these simplifications equation \eqref{eq:btywithc} for $b_{ty}$ is effectively replaced by, 
\begin{equation}\label{eq:btywithc1}
  b_{ty}(r_*, \omega, q)=e^{-i \omega r_*} ~C(\omega,q) \left[1+\frac{i q^2}{\omega} \int_{-\infty}^{r_*} d r_*  f(r_*) \frac{z_H}{z(r_*)} +\mathcal{O}\left(\lambda^2\right) \right]  \, . 
\end{equation}
As it is, the above expression for the $B$--field  has logarithmic divergences near the AdS boundary which we shall address next.

\begin{comment}
\begin{equation} \label{eq:btyfinal1withc1}
   b_{ty} (r_*, \omega, q)=e^{-i \omega r_*} \left[ \frac{1+\frac{i q^2}{\omega} \int_{-\infty}^{r_*} d r_*  f(r_*) \frac{z_H}{z(r_*)} }{1+\frac{i q^2}{\omega} \int_{-\infty}^{0} d r_*  f(r_*) \frac{z_H}{z(r_*)} }+O\left(\lambda^2\right)\right] \,.
\end{equation}
\end{comment}

%Normalization constant $C(\omega,k)$ can be fixed by demanding $\lim_{r_*\to 0} B_{ty}=1$ and we can write $B_{ty}$ as,
%\begin{equation}
   % b_{ty} (r_*, \omega, k)=e^{-i \omega r_*} \left[ \frac{1+  i \omega  \int_{-\infty}^{r_*} d r_* \left(1 - \frac{z_H}{z(r_*)}\right)+ \frac{i q^2}{\omega} \int_{-\infty}^{r_*} d r_*  f(r_*) \frac{z_H}{z(r_*)} }{1+  i \omega  \int_{-\infty}^{z_{\Lambda}} d r_* \left(1 - \frac{z_H}{z(r_*)}\right)+ \frac{i q^2}{\omega} \int_{-\infty}^{z_{\Lambda}} d r_*  f(r_*) \frac{z_H}{z(r_*)} }+O\left(\lambda^2\right)\right]
%\end{equation}
%The integration in the 2nd term turns out to be indeterminate, but we can neglect it by considering small $\omega$ and small $k$ and $\omega$ scales as $\omega \sim k^2$, also $z \sim z_{H}$. So, we will have the following form,
%\begin{equation} \label{btyfinal}
  %  b_{ty} (r_*, \omega, k)=e^{-i \omega r_*} \left[ \frac{1+ \frac{i q^2}{\omega} \int_{-\infty}^{r_*} d r_*  f(r_*) \frac{z_H}{z(r_*)} }{1+\frac{i q^2}{\omega} \int_{-\infty}^{z_{\Lambda}} d r_*  f(r_*) \frac{z_H}{z(r_*)} }+O\left(\lambda^2\right)\right]
%\end{equation}

\subsubsection{Logarithmic divergences} \label{logdivergencebfield}
The 2--form $B$--field has logarithmic divergences while approaching the asymptotic AdS boundary in five spacetime dimensions \cite{Wang:2022xoc,Hofman:2017vwr}. The integral $\displaystyle \int_{-\infty}^{0} d r_*  f(r_*) \frac{z_H}{z(r_*)} $ in \eqref{eq:btywithc1} diverges  logarithmically. This is a generic feature of all form fields and depends on the spacetime dimensions. Similar behavior has been witnessed for the Maxwell gauge field $A_{\mu}$ in asymptotic $\mathrm{AdS}_3$ \cite{Faulkner:2012gt}. To regularize it, we choose a cut off $z_\Lambda$ at the boundary where $\Lambda$ is some energy scale and introduce a counter term in the action. The boundary CFT can be regarded as a matrix--valued field theory, with the current $J$ being a single--trace operator, and then the double trace deformation counter term is\cite{Witten:2001ua},
\begin{equation}\label{eq:counterterm}
\begin{aligned}
    S_{ct} &= \frac{\gamma^2}{2 \kappa(\Lambda)}  \int d^4x J_{\mu \nu} J^{\mu \nu} \,.
\end{aligned}
\end{equation}
This counter term induces a shift in the boundary solution for the $B$--field,
\begin{equation}
\begin{aligned}
\frac{\partial S}{\partial J_{\alpha \beta}} &= \frac{\gamma^2}{2 \kappa(\Lambda)} \frac{\partial {J_{\mu \nu} J^{\mu \nu}}}{\partial J_{\alpha \beta}}    =   \frac{\gamma^2}{ \kappa(\Lambda)} J^{\alpha \beta}  \, .
\end{aligned}
\end{equation}

In the following, only the $B_{ty}$ component is relevant and its boundary value gets a shift, $ \displaystyle \partial S/\partial J_{ty}=  (\gamma^2/\kappa(\Lambda))J^{ty}$. 
On the other hand, using \eqref{eq:Jdef} and 
%\begin{equation}
%    J^{\mu \nu}=-\frac{1}{\gamma^2}\sqrt{-g} H^{r_* \mu \nu},
%\end{equation}
%$H_{r_{*}ty}=\partial_{r_*}B_{ty}-\partial_{t}B_{r_{*}y}$.
substituting $b_{r_{*}y}$ from \eqref{eq:beom1}, we can write $\displaystyle 
    H_{r_{*}ty}= \frac{q^2 f}{(q^2 f -\omega^2)} b_{ty}^{\prime} $, 
which leads to $ \displaystyle J^{ty} %&=-\frac{1}{\gamma^2}\sqrt{-g} g^{r_{*}r_{*}} g^{tt} g^{yy} H_{r_{*} t y}\\&
    =\frac{1}{\gamma^2} \frac{q^2 f} {(q^2 f -\omega^2)} z\,b_{ty}^\prime \simeq \frac{1}{\gamma^2}C(\omega,q) z_{H}\frac{i q^2}{\omega}$ in the hydrodynamic, small $\omega$ limit. 
Incorporating the shift, the near boundary expression for $b_{ty}$ \eqref{eq:btywithc1} is as follows,
\begin{equation}
\begin{aligned}
    b_{ty}(r_*\to 0)&=C(\omega,q) \left[1+ \frac{i q^2}{\omega} z_H \text{ln}z_H - \frac{i q^2}{\omega} z_H \text{ln}z_\Lambda +\frac{1}{\kappa(\Lambda)} z_H \frac{i q^2}{\omega} +\mathcal{O}(\lambda^2)\right]\\&
    %=C(\omega,q) \left[1+ \frac{i q^2}{\omega} z_H \text{ln}z_H - \frac{i q^2}{\omega} z_H \text{ln}\bar{z}\right]\\&
    =C(\omega,q) \left[1+ \frac{i q^2}{\omega} z_H \text{ln}\left(\frac{z_H}{\bar{z}}\right)  +\mathcal{O}(\lambda^2)\right] \,,
\end{aligned}
\end{equation}
where $\displaystyle \bar{z}=z_\Lambda e^{-\frac{1}{\kappa(\Lambda)}}$ is the RG--invariant scale \footnote{The RG scale $\bar{z}$ is the Landau pole of the associated linear response Green's function. It lies between $z_\Lambda$ and the boundary $z=0$ .}. As discussed before, the  normalization constant $C(\omega,q)$ can be fixed by demanding $\lim_{r_*\to 0} b_{ty}=1$, leading up to the final expression for $b_{ty}$ as \footnote{By a slight abuse of notation, $z$ here is the renormalized $z$.},
\begin{equation} \label{eq:btyfinal1}
   b_{ty} (r_*, \omega, q)=e^{-i \omega r_*} \left[ \frac{1+ \frac{i q^2}{\omega} z_H \text{ln}\left(\frac{z_H}{z}\right) }{1+\frac{i q^2}{\omega} z_H \text{ln}\left(\frac{z_H}{\bar{z}}\right) }+\mathcal{O}\left(\lambda^2\right)\right] \,.
\end{equation}
Later in section \ref{sec:pform}, we shall witness power law divergences for specific $p$--forms in AdS--Schwarzschild geometry of various other dimensions. A similar procedure of holographic renormalization applies to those cases, although the expressions for the RG--invariant scale differs. 

\begin{comment}
At any general value of $r_*$, we can write the following expression for $b_{ty}$,
\begin{equation}
 b_{ty} =C(\omega,q) \left[1+ \frac{i q^2}{\omega} z_H \text{ln}\left(\frac{z_H}{z}\right) +O\left(\lambda^2\right)\right].
\end{equation}
Here we abuse notation. Now this $z$ is renormalized $z$.
\end{comment}

\subsubsection{The Diffusion constant}\label{sec:diff}
%Diffusion of conserved $U(1)$ charge is observed here. 
Understanding charge transport is crucial for characterizing the behavior of any system, especially in the presence of a $U(1)$ conservation law. In the dual field theory the charge diffusion constant quantifies how a conserved charge spreads out over time in a system. In this case the diffusion constant $D$ is defined as \footnote{The diffusion constant matches with the one obtained by Nabil \& Diego in \cite{Hofman:2017vwr}.},
\begin{equation}
  D=  \int_{-\infty}^{0} d r_*  f(r_*) \frac{z_H}{z(r_*)}=z_H ~\text{ln}\left(\frac{z_H}{\bar{z}}\right)=\frac{1}{\pi T}~\text{ln}\left(\frac{1}{\pi \bar{z} T}\right)\, ,
\end{equation}
where $T$ is the temperature of the AdS--Schwarzschild black hole. Therefore, the final expression for $b_{ty}$ \eqref{eq:btyfinal1} can be written as,
\begin{equation} \label{eq:btyfinal2}
    b_{ty} (r_*, \omega, q) \sim e^{-i \omega r_*} \left[ \frac{1+ \frac{i q^2}{\omega} z_H ~\text{ln}\left(\frac{z_H}{z}\right) }{1+\frac{i q^2 }{\omega}D} \right]\, .
\end{equation}
The pole at $\omega=-i q^2 D$ in the complex $\omega$--plane leads to a dispersion relation $\omega=-i q^2 D$. Near the horizon, using \eqref{eq:beom1} and \eqref{eq:btyfinal2} we can write $b_{{r_*}y}$ as,
\begin{equation}\label{eq:brybty}
b_{r_*y}=b_{ty}= e^{-i \omega r_*} \left[ \frac{1+ \frac{i q^2}{\omega} z_H ~\text{ln}\left(\frac{z_H}{z}\right) }{1+\frac{i q^2 }{\omega}D} \right]\,.
\end{equation}
In section \ref{sec:2formotoc5dim} we shall use these expressions for computing the $\mathrm{OTOC}$s. The charge diffusion constant $D$ plays a central role in the analysis of the $\mathrm{OTOC}$, connecting the charge transport properties to the approach of scrambled states in the dual field theory. For completeness, and to highlight the differences in the final expressions for $b_{t y}$, we shall study the $B$--fields in seven and six dimensional AdS--Schwarzschild geometry in appendix \ref{sec:b7dim}. %\footnote{The difference is that in six dimensions the $H(r_*)$ vanishes identically.}%.
Similar  computations can be performed for $B$--fields in higher dimensions. No boundary divergences appear for $d>3$. However, lower dimensional cases have power law divergences which are addressed in section \ref{sec:pform}.

\section{$\mathrm{OTOC}$ at late times } \label{sec:2formotoc5dim}
Out-of-time-ordered correlators (OTOCs) are crucial for understanding quantum information dynamics. In the early stages after a perturbation, OTOCs typically exhibit exponential growth, which is indicative of chaotic behavior. This rapid increase reflects how quickly information spreads through the system, a phenomenon often associated with the `butterfly effect' \cite{Shenker:2013pqa}. As time progresses, the behavior of $\mathrm{OTOC}$s transitions to a saturation regime. This late--time behavior indicates that the system has reached a form of equilibrium, where the effects of scrambling have stabilized. In this phase, $\mathrm{OTOC}$s tend to level off, reflecting a balance between chaotic dynamics and thermalization processes.
In this paper we are interested in the late time regime, i.e. for times much larger than the scrambling time. In this regime, the $\mathrm{OTOC}$ between a non--conserved operator and a heavy scalar operator decays exponentially.

We wish to study the impact of a charge conservation law on the $\mathrm{OTOC}$.
In this section we shall discuss about the computation of the $\mathrm{OTOC}$ in scattering approach \cite{Shenker:2014cwa}, i.e. as inner product between form fields in a shock wave background and study its late time behavior. The computations are carried out explicitly for the 2--form $B$--fields but can be extended to any $p$--form in arbitrary dimensions as discussed in section \ref{sec:otherdimotherform}. We shall see that when an $\mathrm{OTOC}$ between a conserved charge operator $J$ and a heavy scalar operator $O$ is evaluated, it exhibits diffusive relaxation rather than the rapid decay typically observed in non--conserved systems. This diffusion arises because the conserved charge density spreads over time, leading to a slower relaxation process.

\subsection{$\mathrm{OTOC}$ as inner product}
The scattering approach \cite{Shenker:2014cwa} to computing $\mathrm{OTOC}$ is reviewed in the section \ref{otocreview}. We can write the boundary $\mathrm{OTOC}$ as an inner product of `in' and `out' asymptotic states, 
\begin{equation} \label{eq:inout2}
\begin{aligned}
| \mathrm{in}\rangle & = J_{0y}^R\left(t_2, x_2\right)O^L\left(t_1, x_1\right)|\text{S--AdS}\rangle \\
| \mathrm{out}\rangle & =O^R\left(t_1, x_1\right)^\dagger J_{0y}^L\left(t_2, x_2\right)^\dagger |\text{S--AdS}\rangle \, ,
\end{aligned}
\end{equation}
where $|\text{S--AdS}\rangle$ denotes the AdS--Schwarzschild black hole thermal double state,
\begin{equation} \label{eq:SAdS}
|\text{S--AdS}\rangle=\sum_n e^{-\frac{\beta}{2} E_n}\left|E_n\right\rangle_L\left|E_n\right\rangle_R \,.
\end{equation}
The superscript $L$ and $R$ on each operator denotes the excitation being created on the left or on the right side, respectively. For example, the insertion of $J^{R/L}_{0y}(t_2)$ at the boundary creates excitation/quanta of the bulk 2--form field $B_{ty}$ near the right/left boundary which then travels towards the blackhole horizon. In the standard low energy supergravity approximation this results in a non trivial profile for the bulk $B$--fields. For the $\mathrm{OTOC}$ computations we shall use the symmetric regularized $\mathrm{OTOC}$ version \cite{Cheng:2021mop,Romero-Bermudez:2019vej,Xu:2022vko}, which can be defined as,
\begin{equation}
\mathcal{C}_{\mathcal{J}, \mathcal{O}}=\operatorname{Tr}\left[\rho^{\frac{1}{2}} J_{0y}^L\left(t_2, x_2\right) O^R\left(t_1, x_1\right) \rho^{\frac{1}{2}} J_{0y}^R\left(t_2, x_2\right) O^L\left(t_1, x_1\right)\right]\,.
\end{equation}
Here, we choose the scalar operators to have large conformal dimensions $\Delta$ and hence we can write the correlator $\mathcal{C}_{\mathcal{J}, \mathcal{O}}$ as an inner product between two $J$ insertions. This is equivalent to calculating the inner product between two 2--form fields with and without the presence of a shockwave, to be discussed below \footnote{On--shell variation of the action yields a boundary term associated with the symplectic form, $\displaystyle \delta S\bigr|_{\mathrm{on-shell}}=\int \delta \mathcal{L}=\int \delta \phi \frac{\delta \mathcal{L}}{\delta \phi}+\int d (\pi(x)\delta\phi(x))=\int \pi(x)\delta\phi(x)\bigr|_{\mathrm{boundary}}=\tilde{\Theta}$ . This boundary term defines the pre--symplectic structure which when pulled back to the on--shell condition, gives the symplectic form, $\displaystyle \Theta=\int \delta \pi(x) \wedge \delta \phi(x)\bigr|_{\mathrm{boundary}}$ . The symplectic term naturally comes as a gauge invariant inner product for quadratic Lagrangians like \eqref{eq:bfieldaction}.}. 
\begin{figure}[h!]
        \centering
\tikzset{every picture/.style={line width=.75pt}} %set default line width to 0.75pt        

\begin{tikzpicture}[x=0.75pt,y=0.75pt,yscale=-1,xscale=1]
\path (0,200); 
%set diagram left start at 0, and has height of 300

%Straight Lines [id:da7229253864197858] 
\draw    (200,37) -- (201,210) ;
%Straight Lines [id:da9347370150642308] 
\draw    (360,30) -- (360,210) ;
%Straight Lines [id:da18762923001273868] 
\draw    (360,30) -- (201,210) ;
%Straight Lines [id:da4427847417709402] 
\draw    (360,210) -- (200,37) ;
%Shape: Free Drawing [id:dp48449251331879983] 
\draw  [color={rgb, 255:red, 208; green, 2; blue, 27 }  ,draw opacity=1 ][line width=0.75] [line join = round][line cap = round] (201,38) .. controls (201,29) and (213.59,40.38) .. (216,39) .. controls (218.98,37.3) and (222.36,31.8) .. (225,34) .. controls (226.81,35.51) and (228.21,40.53) .. (230,39) .. controls (232.15,37.16) and (233.85,34.84) .. (236,33) .. controls (237.16,32) and (243.23,39.41) .. (247,38) .. controls (248.03,37.61) and (252.42,30.42) .. (254,32) .. controls (256.85,34.85) and (256.36,39.11) .. (263,38) .. controls (266.4,37.43) and (269.83,26.42) .. (274,30) .. controls (276.56,32.2) and (277.27,38.18) .. (282,37) .. controls (286.27,35.93) and (285.46,28.88) .. (289,28) .. controls (295,26.5) and (295.13,37.55) .. (299,36) .. controls (303.95,34.02) and (306.76,29.82) .. (311,27) .. controls (315.01,24.33) and (315.08,37.94) .. (319,35) .. controls (321.76,32.93) and (322.12,27.37) .. (327,29) .. controls (329.31,29.77) and (332.97,37.03) .. (335,35) .. controls (337.03,32.97) and (337.97,30.03) .. (340,28) .. controls (341.79,26.21) and (344.8,33.12) .. (347,34) .. controls (350.19,35.28) and (352.66,29.96) .. (354,29) .. controls (355.79,27.72) and (357.87,31) .. (360,31) ;
%Shape: Free Drawing [id:dp05011670582074512] 
\draw  [color={rgb, 255:red, 208; green, 2; blue, 27 }  ,draw opacity=1 ][line width=0.75] [line join = round][line cap = round] (203,208) .. controls (202.64,208) and (206.09,213.17) .. (210,212) .. controls (213.86,210.84) and (217.4,205.2) .. (221,207) .. controls (223.53,208.26) and (222.59,213.8) .. (228,212) .. controls (230.85,211.05) and (232.48,203.65) .. (236,206) .. controls (239.81,208.54) and (241.26,214.74) .. (247,209) .. controls (248.84,207.16) and (249.67,201.84) .. (252,203) .. controls (255.7,204.85) and (257.87,211.28) .. (263,210) .. controls (268.11,208.72) and (268.25,201.38) .. (273,199) .. controls (278.41,196.29) and (284.32,209.68) .. (288,206) .. controls (290.22,203.78) and (290.19,199.41) .. (293,198) .. controls (295.95,196.52) and (297.52,202.83) .. (300,205) .. controls (303.7,208.24) and (307.24,199.69) .. (310,199) .. controls (314.08,197.98) and (313.63,202.47) .. (317,205) .. controls (320.6,207.7) and (325.74,197.87) .. (330,200) .. controls (333.02,201.51) and (332.03,206.01) .. (335,207) .. controls (339.9,208.63) and (341.27,202.9) .. (346,205) .. controls (347.95,205.87) and (348.96,208.37) .. (351,209) .. controls (353.57,209.79) and (359,207.31) .. (359,210) ;
%Straight Lines [id:da7348865329411891] 
\draw    (247,139) -- (201,190) ;
%Straight Lines [id:da058301336022897] 
\draw    (200,57) -- (235,97) ;
%Straight Lines [id:da06455753400033748] 
\draw    (360,50) -- (318,98) ;
%Straight Lines [id:da5575885976702107] 
\draw    (360,192) -- (319,147) ;
%Shape: Free Drawing [id:dp8924684114414316] 
\draw  [color={rgb, 255:red, 0; green, 0; blue, 0 }  ,draw opacity=1 ][line width=0.75] [line join = round][line cap = round] (332,164) .. controls (332,165.33) and (332,169.33) .. (332,168) .. controls (332,166) and (331.37,163.9) .. (332,162) .. controls (332.21,161.37) and (333.4,161.7) .. (334,162) .. controls (335.61,162.8) and (336.88,164) .. (340,164) ;
%Shape: Free Drawing [id:dp41784574210540326] 
\draw  [color={rgb, 255:red, 0; green, 0; blue, 0 }  ,draw opacity=1 ][line width=0.75] [line join = round][line cap = round] (337,68) .. controls (339.49,68) and (341.82,66.55) .. (344,66) .. controls (345.02,65.74) and (347,63.95) .. (347,65) .. controls (347,67.56) and (345,74.5) .. (345,74) ;
%Shape: Free Drawing [id:dp34955119910669696] 
\draw  [color={rgb, 255:red, 0; green, 0; blue, 0 }  ,draw opacity=1 ][line width=0.75] [line join = round][line cap = round] (218,79) .. controls (218,81) and (218,87) .. (218,85) .. controls (218,82.67) and (216.24,79.54) .. (218,78) .. controls (220.01,76.24) and (223.33,78) .. (226,78) ;
%Shape: Free Drawing [id:dp19864934648855326] 
\draw  [color={rgb, 255:red, 0; green, 0; blue, 0 }  ,draw opacity=1 ][line width=0.75] [line join = round][line cap = round] (537,150) .. controls (537,151) and (537,152) .. (537,153) ;
%Shape: Free Drawing [id:dp4875677785608763] 
\draw  [color={rgb, 255:red, 0; green, 0; blue, 0 }  ,draw opacity=1 ][line width=0.75] [line join = round][line cap = round] (230,154) .. controls (230,154) and (230,154) .. (230,154) ;
%Shape: Free Drawing [id:dp45470016726938545] 
\draw  [color={rgb, 255:red, 0; green, 0; blue, 0 }  ,draw opacity=1 ][line width=0.75] [line join = round][line cap = round] (223,156) .. controls (227,156) and (229.41,154) .. (234,154) .. controls (234.33,154) and (233.11,153.68) .. (233,154) .. controls (232.19,156.43) and (232,160.49) .. (232,163) ;

% Text Node
\draw (369,43) node [anchor=north west][inner sep=0.75pt]   [align=left] {$O^R(t_1)$};
% Text Node
\draw (274,46) node [anchor=north west][inner sep=0.75pt]   [align=left] {$|\mathrm{out}\rangle$};
% Text Node
\draw (372,181) node [anchor=north west][inner sep=0.75pt]   [align=left] {$J^R_{0y}(t_2)$};
% Text Node
\draw (145,46) node [anchor=north west][inner sep=0.75pt]   [align=left] {$J^L_{0y}(t_2)$};
% Text Node
\draw (143,182) node [anchor=north west][inner sep=0.75pt]   [align=left] {$O^L(t_1)$};
% Text Node
\draw (274,180) node [anchor=north west][inner sep=0.75pt]   [align=left] {$|\mathrm{in}\rangle$};

\end{tikzpicture}
\caption{The `in' and `out' states in the Penrose diagram with scalar and current operators}  
\end{figure}
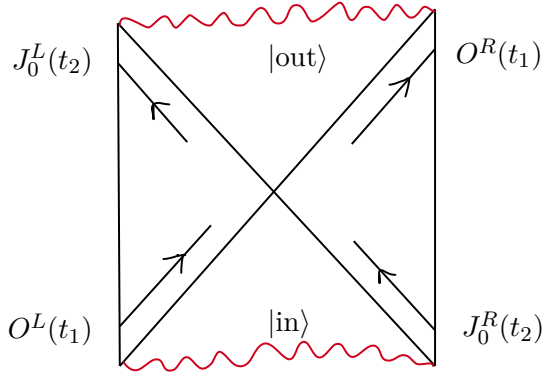

\subsection{The inner product of two $B$--fields}
As we are working with gauge fields in a curved geometry, a gauge invariant inner product between two 2--form fields $B_1$ and $B_2$ can be defined as,
\begin{equation}
\left(B_1, B_2\right)=\int \sqrt{h} n^\mu\left((H_1)_{\mu \nu \rho}^* B_2^{\nu \rho}-(H_2)_{\mu \nu \rho} B_1^{* \nu \rho}\right) \, ,
\end{equation}
where we choose to integrate on the constant $u \sim 0$ slice with $h_{\mu \nu}$ as the induced metric on the hypersurface and $n^{\mu}$ being the unit vector normal to this hypersurface \cite{Cheng:2021mop}. Here, the $u$ and $v$ are related to the usual $t$ and $r_*$ coordinates by the transformations,
\begin{equation}
u =-e^{\frac{2 \pi}{\beta}(r_*\,-t)}\,, \quad 
v  =e^{\frac{2 \pi}{\beta}(t+r_*)} \quad \text{and} \quad 
\left(\frac{\beta}{2 \pi}\right)^2 g^{u v}  =2 u v g^{t t} \,.
\end{equation}

We shall now specialize to $B$--fields in the five dimensional AdS--Schwarzschild geometry. The inner product takes the form,
\begin{equation}\label{eq:innerproduct}
\begin{aligned}
\left(B_1, B_2\right)%&=\int d v d^3 x ~\sqrt{h} n^v\left(H_{1 v \nu \rho}^* B_2^{\nu \rho}-H_{2 v \nu \rho} B_1^{* \nu \rho}\right)\\&
%=\int d v d^3 x ~\sqrt{h} n^v\left(H_{1 v \nu \rho}^* B_2^{\nu \rho}-H_{2 v \nu \rho} B_1^{* \nu \rho}\right)\\&
&=\int d v d^3 \vec{x} ~\sqrt{h} n^v g^{yy} g^{vu}\left((H_1)_{vuy}^* (B_2)_{vy}-(H_2)_{vuy} (B_1)_{vy}^*\right)\\
&=\int d v d^3 \vec{x} ~\sqrt{h} n^v g^{yy} g^{tt}\left((H_1)_{r_*ty}^* (B_2)_{vy}-(H_2)_{r_*ty} (B_1)_{vy}^*\right)\,,
\end{aligned}
\end{equation}
where $\displaystyle \sqrt{h}=1/z^3 $ and $ \displaystyle n_v=\partial/\partial v$ is the unit normal vector. The fields $b_{vy}$ and $H_{vuy}$ can be written in terms of $b_{r_*y}$ and $b_{ty}$ and they are related by the following transformations,
\begin{equation}\label{eq:hvuy}
    b_{vy} =\frac{1}{v} \frac{\beta}{2 \pi} b_{ty}\, \quad \text{and} \quad 
    H_{vuy}=\left(\frac{\beta}{2 \pi}\right)^2 \frac{1}{2 u v} H_{r_*ty} \,.
\end{equation}
In the real space, we can write $B_{vy}$ as,
\begin{equation}\label{eq:bshi}
\begin{aligned}
    B_{vy}=&\int d\omega d^3\vec{q}~\frac{1}{v} \frac{\beta}{2 \pi}  e^{-i \omega r_*} \left[ \frac{1+ \frac{i q^2}{\omega} z_H ~\text{ln}\left(\frac{z_H}{z}\right) }{1+\frac{i q^2 }{\omega}D} \right] e^{-i \omega t+ i \vec{q}\cdot \vec{x}}\\=
    &\partial_v\left(\int d \omega d^3 \vec{q} ~\frac{i}{\omega}\left[ \frac{1+ \frac{i q^2}{\omega} z_H ~\text{ln}\left(\frac{z_H}{z}\right) }{1+\frac{i q^2 }{\omega}D} \right]  v^{-i \frac{\beta \omega}{2 \pi}} e^{i \vec{q}\cdot \vec{x}}\right)=\partial_v \psi(v, \vec{x}) \, ,
\end{aligned}
\end{equation}
where $\psi(v,\Vec{x})$ is,
\begin{equation}
    \psi(v,\Vec{x})=\int d \omega d^3 \vec{q} ~\frac{i}{\omega}\left[ \frac{1+ \frac{i q^2}{\omega} z_H ~\text{ln}\left(\frac{z_H}{z}\right) }{1+\frac{i q^2 }{\omega}D} \right]  v^{-i \frac{\beta \omega}{2 \pi}} e^{i \vec{q}\cdot \vec{x}}\, .
\end{equation}
Similarly, in the real space $H_{vuy}$ as in \eqref{eq:hvuy} can be written as,
\begin{equation}
    \begin{aligned}
        g_{yy}g^{tt}H_{r_*ty}= &\int d \omega d^3 \vec{q} ~\frac{iq^2}{\omega}\left[ \frac{1+ \frac{i q^2}{\omega} z_H ~\text{ln}\left(\frac{z_H}{z}\right) }{1+\frac{i q^2 }{\omega}D} \right]  v^{-i \frac{\beta \omega}{2 \pi}} e^{i \vec{q}\cdot \vec{x}}=\nabla^2 \psi(v, \vec{x})\, .
    \end{aligned}
\end{equation}
Finally, we can write the inner product  \eqref{eq:innerproduct} as,
\begin{align}\label{eq:innerbb}
\left(B_1, B_2\right) & =\int d v d^3 \vec{x} ~\sqrt{h}g^{yy}g^{yy}\left[\nabla^2 \psi_1^*(v, \vec{x}) \partial_v \psi_2(v, \vec{x})-\nabla^2 \psi_2(v, \vec{x}) \partial_v \psi_1^*(v, \vec{x})\right] \\
& =\int d v d^3 \vec{x} ~\sqrt{h}g^{yy}g^{yy} \left[\vec{\nabla} \psi_2(v, \vec{x}) \cdot \partial_v \vec{\nabla} \psi_1^*(v, \vec{x})-\vec{\nabla} \psi_1^*(v, \vec{x}) \cdot \partial_v \vec{\nabla} \psi_2(v, \vec{x})\right] \, . \nonumber
\end{align}
We shall now use this form of the inner product in the $\mathrm{OTOC}$ computations.

\subsection{Shock waves and the scattering states}
%and equation \eqref{eq:innerbb} can be written as, \AC{Check this? Isn't $\sqrt{h}g^{xx}g^{xx} $ function of $v$, factors of $2 \pi$ etc.} \AC{Where exactly have we used this?}
%\begin{equation}\label{otocmomentum}
% \left(B_1, B_2\right)=i \int d p d^3 x ~\sqrt{h}g^{xx}g^{xx} ~(2p)\left[\vec{\nabla} \psi_1^*(p, \vec{x}) \cdot  \vec{\nabla} \psi_2(p, \vec{x})\right] \, .
%\end{equation}
%\begin{equation}\label{otocmomentum}
%\begin{aligned}
%    \left(B_1, B_2\right) &=\int d v d^3 x ~\sqrt{h}g^{xx}g^{xx} \left[-\vec{\nabla} \psi_1^*(v, \vec{x}) \cdot (-iq) \vec{\nabla} \psi_2(v, \vec{x})+\vec{\nabla} \psi_2(v, \vec{x}) \cdot (iq) \vec{\nabla} \psi_1^*(v, \vec{x})\right]\\
%    &= \int d v d^3 x ~\sqrt{h}g^{xx}g^{xx} ~(2iq)\left[\vec{\nabla} \psi_1^*(v, \vec{x}) \cdot  \vec{\nabla} \psi_2(v, \vec{x})\right]
%\end{aligned}
%\end{equation}
The expressions for $\psi$ and $\psi^*$ in the backdrop of the asymptotic `in' and `out' states as discussed in \eqref{eq:inout2} and in \eqref{eq:SAdS} can be found from the left$(L)$ and right$(R)$ $B$--fields in the AdS--Schwarzschild thermal background with inverse temperature $\beta$. In general, the $B$--field is a linear superposition of the in--falling and the out--going solutions to the equation of motion, and depends on the particulars of state construction of interest. Following \cite{Cheng:2021mop}, we choose the coefficients such that the field has positive Kruskal frequency for the in--falling mode and negative Kruskal frequency for the out--going mode. For this purpose, we shall use the following combination for the $b_{\mu y}^R(r, \omega, q)$ (an analogous expression can be written down for $b_{\mu y}^L(r, \omega, q)$), 
\begin{equation}
b_{\mu y}^R(r, \omega, q)=(1+n(\omega)) b_{{\mu y}_{\text {in--falling }}}^R(r, \omega, q)-n(\omega) b^{R}_{{\mu y}_{\text {out--going }}}(r, \omega, q)\, ,
\end{equation}
 where $\displaystyle n(\omega)=\frac{1}{e^{\beta \omega}-1}$ is  the Boltzmann factor. From equation \eqref{eq:brybty} we get the dispersion relation,  $\omega=-i D q^2$. Note that for small values of $q$, we have  $\displaystyle (1+n(\omega) )  \sim \frac{1}{2 \sin{\left(\frac{\beta D q^2}{2}\right)}}~$.

\begin{comment}

\subsubsection{Scattering States}
We consider the following correlator,

\begin{equation}
\mathcal{C}_{\mathcal{J}, \mathcal{O}}=\operatorname{Tr}\left[\rho^{\frac{1}{2}} J_0\left(t_2, \vec{x}_2\right) O\left(t_1, \vec{x}_1\right) \rho^{\frac{1}{2}} J_0\left(t_2, \vec{x}_2\right) O\left(t_1, \vec{x}_1\right)\right]
\end{equation}

It has a representation as an inner product of $in$ and $out$ states.
\begin{equation}
\begin{aligned}
\mid \textit {in}\rangle & =O^L\left(t_1, \vec{x}_1\right) J_0^R\left(t_2, \vec{x}_2\right)|S-A d S\rangle \\
\mid \textit {out}\rangle & =J_0^L\left(t_2, \vec{x}_2\right) O^R\left(t_1, \vec{x}_1\right)|S-A d S\rangle
\end{aligned}
\end{equation}

where $|S-A d S\rangle$ denotes the Schwarzschild-AdS black hole thermal double state,
\begin{equation}
|S-A d S\rangle=\sum_n e^{-\frac{\beta}{2} E_n}\left|E_n\right\rangle_L\left|E_n\right\rangle_R .
\end{equation}

The superscript $L$ and $R$ on each operator means that the excitation is created on the left or right side, respectively. For example, $J_0^R\left(t_2\right)$ creates a $B$-field on the right. We denote the corresponding gauge field by $B_{\mu y}^R$.

\end{comment}
For the in--falling part, we shall almost use the expression as given in \eqref{eq:bshi}, except for inserting an extra factor of 1 in the denominator. This change doesn't affect the late time behavior, but simplifies the analysis,
\begin{equation}
\begin{aligned}
B_{v y_{\text {in--falling }}}^R & =\partial_v \psi_{\text {in--falling }}^R(v, \vec{x})\, , \\
\psi_{\text {in--falling }}^R(v, \vec{x})  &=-i\int  \frac{d^3 \vec{q}}{(2 \pi)^3}  \frac{\frac{z_H}{D} \text{ln}\left(\frac{z_H}{z}\right)}{\left(v e^{-\frac{2 \pi}{\beta} t_2}-1\right)^{\frac{\beta D q^2}{2 \pi}}} \theta\left(v-e^{\frac{2 \pi}{\beta} t_2}\right) e^{i \vec{q} \cdot\left(\vec{x}-\vec{x}_2\right)} \\ & \quad + i \int \frac{d^3 \vec{q}}{(2 \pi)^3} \frac{1}{\left(v e^{-\frac{2 \pi}{\beta} t_2}-1\right)^{\frac{\beta D q^2}{2 \pi}}}\theta\left(v-e^{\frac{2 \pi}{\beta} t_2}\right) e^{i \vec{q} \cdot\left(\vec{x}-\vec{x}_2\right)} \,.
\end{aligned}
\end{equation}
The out--going part of the $B$--field is the complex conjugate of the in--falling part and hence it is proportional to $u$. As we are evaluating the inner product on the $u \sim 0$ slice for the $\mathrm{OTOC}$ computations we can neglect the outgoing part's contribution. Incorporating the Boltzmann factor near the diffusive pole, we choose the final ansatz for $B_{vy}^R$ as \footnote{Real space form of $\psi^R$ and $\psi_{\text{in--falling}}^R$ are related by analytic continuation \cite{Cheng:2021mop}.},
\begin{equation}
\begin{aligned}
B_{v y}^R & =\partial_v \psi^{R}(v, \vec{x})\,, \\
\psi^R(v, \vec{x}) & =-i\int \frac{d^3 \vec{q}}{(2 \pi)^3}  \frac{\frac{z_H}{D} \text{ln}\left(\frac{z_H}{z}\right)}{2 \sin{\left(\frac{\beta D q^2}{2}\right)} \left(1-v e^{-\frac{2 \pi}{\beta} t_2}\right)^{\frac{\beta D q^2}{2 \pi}}} \,e^{i \vec{q} \cdot\left(\vec{x}-\vec{x}_2\right)} \\ & \qquad + i \int \frac{d^3 \vec{q}}{(2 \pi)^3} \frac{1}{2 \sin{\left(\frac{\beta D q^2}{2}\right)} \left(1-v e^{-\frac{2 \pi}{\beta} t_2}\right)^{\frac{\beta D q^2}{2 \pi}}} \,e^{i \vec{q} \cdot\left(\vec{x}-\vec{x}_2\right)}\,  .
\end{aligned}
\end{equation}
By symmetry, the left $B$--field  $B_{vy}^L$ is related to the right $B$--field $B_{vy}^R$ via reflection in the  $(u,v)$ coordinates i.e., $(u,v) \rightarrow (-u,-v)$.

\begin{comment}
We can write the following expression for the left part,
\begin{equation}
\begin{aligned}
B_{v y}^L & =\partial_v \psi^{L}(v, \vec{x}), \\
\psi^L(v, \vec{x}) & =\int \frac{d^3 q}{(2 \pi)^3}  \frac{ i \frac{z_H}{D} \text{ln}\left(\frac{z_H}{z}\right)}{\left(1+v e^{-\frac{2 \pi}{\beta} t_2}\right)^{\frac{\beta D q^2}{2 \pi}}} \theta\left(v-e^{\frac{2 \pi}{\beta} t_2}\right) e^{i \vec{q} \cdot\left(\vec{x}-\overrightarrow{x_2}\right)} \\ & - \int \frac{d^3 q}{(2 \pi)^3} \frac{i}{\left(1+v e^{-\frac{2 \pi}{\beta} t_2}\right)^{\frac{\beta D q^2}{2 \pi}}}\theta\left(v-e^{\frac{2 \pi}{\beta} t_2}\right) e^{i \vec{q} \cdot\left(\vec{x}-\overrightarrow{x_2}\right)} .
\end{aligned}
\end{equation}
\end{comment}

%The scalar operator $O^R$ and $O^L$ create scalar mode in the bulk. For simplicity, we assume that the scalar operator $O$ has a large conformal dimension, corresponding to a particle in bulk with a large mass. As a consequence, we can treat the scalar particle semiclassically. As it moves deep into the bulk, the scalar mode carries the shockwave along with it, which modifies the $B$-field's wave function as we will analyze in the next section. In contrast, we will neglect the back-reaction from the $B$-field on this scalar mode.

We are interested in the large time limit of the $\mathrm{OTOC}$ and shall take the limits $t_1 \gg \beta$ and $t_2 \ll-\beta$, which means that the bulk geodesic of the scalar particle created by the scalar operators $O^R$ and $O^L$ hovers around $u=\epsilon \sim 0$ slice. Further we shall assume a semi--classical treatment of the scalar particle owing to their large mass (the scalar operators have large conformal dimensions). As it moves deep into the bulk, the scalar mode carries a shockwave along with it, which in turn modifies the $B$--field's wave function but we shall neglect the backreaction from the $B$--field itself. The shockwave geometry is described by a metric that contains a singularity near the horizon,
\begin{equation}
d s^2=2 g_{u v} d u[d v-\delta(u) h(\vec{x}) d u]+g_{x x} d \Omega_d\,,
\end{equation}
where $\displaystyle h(\vec{x}) \sim \frac{\Delta_O}{N} e^{\frac{2 \pi}{\beta} t_1-\mu\left|\vec{x}-\vec{x}_1
\right|}$ 
%\AC{How to compare this with what we have written in the introduction? Also, check the calculations in section 2.4 of Swingle. I think there are typos} %
and $\displaystyle \mu=\sqrt{\frac{2 d}{d+1}} \frac{2 \pi}{\beta}$ with $d=3$ (five dimensional geometry) \footnote{For $h(\vec{x})$, $N$ is the number of d.o.f. and we neglected the power law fall--off in the transverse directions in favor of the exponential fall-off.}. The shockwave shifts the $v$ coordinate by an amount $h$ and the expressions for the left and the right  side fields take the form \cite{Cheng:2021mop,Shenker:2014cwa},
\begin{equation}
    \begin{split}
        \psi_1(v,\vec{x})=\psi^{L}(v,\vec{x})\, ,\\
        \psi_2(v,\vec{x})=\psi^{R}(v-h(\vec{x}),\vec{x})\, .
    \end{split}
\end{equation}
The final piece that remains is to compute the inner product \eqref{eq:innerbb}, which will be equal to the $\mathrm{OTOC}$.

\subsection{The $\mathrm{OTOC}$}
It is easier to compute $\mathrm{OTOC}$s in momentum space, where $\displaystyle  \psi(p,\vec{x})=\int d v ~e^{ipv} \psi(v,\vec{x})~$. Substituting $\psi_1$ and $\psi_2$ in \eqref{eq:innerbb}, we first calculate $\vec{\nabla}\psi$'s in momentum space. 
\begin{align}
\nonumber \vec{\nabla} \psi_1(p, \vec{x}) & =\int \frac{d^3 \vec{q}_1}{(2 \pi)^3} \frac{2 \pi \vec{q_1} \left( \frac{z_H}{D} \text{ln}\frac{z_H}{z}-1\right)}{2 \sin \left(\frac{\beta D q_1^2}{2}\right)} \frac{p^{\frac{\beta D q_1^2}{2 \pi}-1} e^{D q_1^2 t_2}}{\Gamma\left(\frac{\beta D q_1^2}{2 \pi} \right)} \,e^{i p e^{\frac{2 \pi}{\beta} t_2}} e^{-\frac{i}{4}\beta D q_1^2} e^{i \vec{q_1} \cdot\left(\vec{x}-\vec{x_2}\right)}  \theta(p)\,, \\ 
\vec{\nabla} \psi_2(p, \vec{x}) & =\int \frac{d^3 \vec{q}_2}{(2 \pi)^3} \frac{2 \pi \vec{q_2} \left( \frac{z_H}{D} \text{ln}\frac{z_H}{z}-1\right)}{2 \sin \left(\frac{\beta D q_2^2}{2}\right)} \frac{p^{\frac{\beta D q_2^2}{2 \pi}-1} e^{D q_2^2 t_2}}{\Gamma\left(\frac{\beta D q_2^2}{2 \pi} \right)} \\ \nonumber
& \qquad \times e^{-i p e^{\frac{2 \pi}{\beta} t_2}} e^{\frac{i}{4}\beta D q_2^2} e^{i p\, h(\vec{x})} e^{i \vec{q_2} \cdot\left(\vec{x}-\vec{x}_2\right)} \theta(p) \, .
\end{align}
Note that if we only keep the leading order terms in $q^2$, then the term $\displaystyle \sin \left(\beta D q^2/2\right)$ cancels $\displaystyle \Gamma\left(\frac{\beta}{2 \pi} D q^2\right)$. As long as we are considering large transverse coordinate separation, this approximation should be qualitatively correct. Plugging these into equation \eqref{eq:innerbb} and approximating $z/z_{H} \sim 1$, we obtain,
\begin{align}
\left(B_1, B_2\right) & \sim \int d p \int d^3 \vec{x} \int \frac{d^3 \vec{q}_1}{(2 \pi)^3} \int \frac{d^3 \vec{q}_2}{(2 \pi)^3}(\vec{q_1} \cdot \vec{q_2}) ~z\left( \frac{z_H}{D} \ln \frac{z_H}{z}-1\right)^2 \\
\nonumber & \qquad \times  e^{D (q_1^2 +q_2^2) t_2} p^{\frac{\beta D (q_1^2+q_2^2)}{2 \pi}-1} e^{-2 i p \,e^{\frac{2\pi}{\beta}t_2}} e^{i p\, h\left(\vec{x}\right)} e^{-i\left(\vec{q}_1-\vec{q}_2\right) \cdot\left(\vec{x}-\vec{x}_2\right)} e^{\frac{i}{4}\beta D (q_1^2+q_2^2)} \, .
\end{align}
%where $\displaystyle h(\vec{x})=\frac{\Delta_O}{N} e^{\frac{2 \pi}{\beta} t_1-\mu\left|\vec{x}-\vec{x}_1\right|}$ and $\mu=\sqrt{\frac{2 d}{d+1}} \frac{2 \pi}{\beta}$ with $d=3$.\\
Translating time $t\rightarrow t-t_2$ and integrating over $p$, we get the following expression,
\begin{equation}
\begin{aligned}
 \left(B_1, B_2\right) \sim & ~z\left( \frac{z_H}{D} \ln \frac{z_H}{z}-1\right)^2 \int d^3 \vec{x} \int \frac{d^3 \vec{q}_1}{(2 \pi)^3} \int \frac{d^3 \vec{q}_2}{(2 \pi)^3}\left(\vec{q_1} \cdot \vec{q_2}\right)\\
&\quad \times \Gamma\left[D\left(q_1^2+q_2^2\right)\right]\left[2+\frac{{\Delta}_{O}}{N} e^{\frac{2 \pi}{\beta}t_{12}-\mu\left|\vec{x}-\vec{x}_{12}\right|}\right]^{-D\left(q_1^2+q_2^2\right)} e^{-i\left(\vec{q}_1-\vec{q}_2\right) \cdot \vec{x}} \,.
\end{aligned}
\end{equation}
To perform the integrals we shall keep only the leading order dependence in $q$ for the Gamma functions. We change the integration variables to $\displaystyle \vec{K}=\frac{\vec{q}_1+\vec{q}_2}{2}$ and $\displaystyle \vec{\kappa}=\frac{\vec{q}_1-\vec{q}_2}{2}$, this gives 
\begin{equation}\label{eq:bbsemifinal}
\begin{aligned}
\left(B_1, B_2\right) & \sim z\left( \frac{z_H}{D} \ln \frac{z_H}{z}-1\right)^2 \left (\int d^3 \vec{x} \int \frac{d^3 \vec{K}}{(2 \pi)^3} \int \frac{d^3 \vec{\kappa}}{(2 \pi)^3}  \frac{8(K^2-\kappa^2)}{D\left(K^2+\kappa^2\right)} \right.\\
& \qquad \left. \times \left[2+\frac{\Delta_O}{N} e^{t_{12}-\mu\left|\vec{x}-\vec{x}_{12}\right|}\right]^{-2 D\left(K^2+\kappa^2\right)}  e^{-2 i \vec{\kappa} \cdot \vec{x}}\right) \\
& =z\left( \frac{z_H}{D} \ln \frac{z_H}{z}-1\right)^2 \int \frac{d^3 \vec{x}}{(4 \pi)^3}\left[\frac{3}{D^{4}} E\left(\ln g, \frac{|\vec{x}|^2}{2 D}\right)-\frac{1}{2 D^{4}} \frac{1}{(\ln g)^3} e^{-\frac{|\vec{x}|^2}{2 D \ln g}}\right]\,,
\end{aligned}
\end{equation}
where the functions $g$ and $E$ are defined as
\begin{equation}
g  =2+\frac{\Delta_O}{N} e^{t_{12}-\mu\left|\vec{x}-\vec{x}_{12}\right|} \quad \text{and} \quad 
E(z, a)  =\int_z^{\infty} d y \frac{1}{y^{4}} e^{-\frac{a}{y}} \, .
\end{equation}
Since both terms in \eqref{eq:bbsemifinal} contain the factor $e^{-\frac{|\vec{x}|^2}{2 D \ln g}}$, we expect the integral to receive its dominant contribution from $|\vec{x}| \sim 0\,$. Integrating $\vec{x}$ over this saddle point gives a factor of $\displaystyle (2 D \ln g)^{\frac{3}{2}}$. Therefore, we finally obtain
\begin{equation}
\mathrm{OTOC} = \left(B_1, B_2\right) \sim \frac{z \left( \frac{z_H}{D} \ln \frac{z_H}{z}-1\right)^2}{ D^{\frac{3}{2}+1}\left[\ln \left(2+\frac{\Delta_O}{N} e^{\frac{2 \pi}{\beta} t_{12}-\mu\left|\vec{x}_{12}\right|}\right)\right]^{\frac{3}{2}}}\,  .
\end{equation}

At early time, this expression admits a large $N$ expansion in which the leading term still grows exponentially with time. However, in the late time limit, the $\log$ of the exponentially growing part in the denominator gives rise to a power law time decay behavior, $\mathrm{OTOC} \sim t^{-\frac{3}{2}}$ \footnote{In case of reduced transverse spherical symmetry as discussed earlier the falloff is $\mathrm{OTOC} \sim t^{-1}$ .}. Hence, we find significant difference from the $\mathrm{OTOC}$ of non--conserved operators in the late time regime. This occurs due to the
hydrodynamical property of the conserved current. The particles sourced by these conserved operators in the bulk spread over a large region of space-time leading to this kind of diffusive power law behavior at late times. This type of power law behavior is already known for random circuit models \cite{Khemani:2017nda,Nahum_2018,Rakovszky_2018} and for certain holographic theories \cite{Cheng:2021mop}.

\section{Higher--form fields and other dimensions}\label{sec:otherdimotherform}
Higher--form fields in asymptotic AdS geometries offer a compelling avenue for exploring fundamental questions in theoretical physics. Their unique properties provide insights into both gravitational dynamics and quantum field theory through holographic dualities, e.g. higher--form global symmetry at the boundary of AdS are related to the higher--form fields in the bulk of AdS \cite{Gomes_2023,Hofman:2017vwr}. Till now we have discussed the 2--form field, its dynamics and the late time behavior of its $\mathrm{OTOC}$.
Now we shall generalize the discussions of the previous two sections to $p$--form fields in $d+2$ dimensional ($d \geq 2$) AdS--Schwarzschild black hole geometries and discuss the late time behavior of their respective $\mathrm{OTOC}$s \footnote{We note that in the case of gauge fields $A_\mu$ in $\mathrm{AdS}_3$, the anticipation of a universal relation between the long--distance transport coefficients (thermal) and the parameters  describing the short--distance singularities of the current--current correlator is true. The reason is that in 2d CFTs, the conformal group is large enough and the vacuum state is related to the thermal state by a symmetry transformation. In this case, the two point current--current correlator has no diffusive poles and shows no hydrodynamical modes as discussed in detail in \cite{Kovtun:2008kx,Faulkner:2012gt}. We thank the referee for bringing this to our notice.}. We have also highlighted a specific form field (3--form) to discuss about the power law divergences at the boundary and its regularization. 

\subsection*{A $p$--form field in $(d+2)$ dimensional AdS--Schwarzschild geometry }\label{sec:pform}
The action of a minimally coupled $p$--form field $P$ can be written as
\begin{equation}\label{eq:actionpform}
S[P]=-\frac{1}{2(p+1)} \int \mathrm{d}^{d+2} x \sqrt{-g} (\mathrm{d} P)^2 \, ,   
\end{equation}
where $\mathrm{d} P$ is the $(p+1)$--form field strength. The theory has a $\mathrm{U}(1)$ gauge symmetry, i.e. $P \rightarrow P+\mathrm{d} \Lambda$ for an arbitrary $(p-1)$--form $\Lambda$. The equation of motion for the above action are
\begin{equation}
 \partial_{\mu_1}\left(\sqrt{-g}  ~\mathrm{d} P^{\mu_1 \cdots \mu_p}\right)=0   \, .
\end{equation}
As in equation \eqref{eq:Jdef} for the 2--form field, we can define the boundary current operator as,
\begin{equation}\label{eq:Jdefpform}
    J^{\mu_1 \mu_2 \cdots \mu_p}=-\frac{1}{\gamma^2}\sqrt{-g} H^{r_{*} \mu_1 \mu_2 \cdots \mu_p}\, .
\end{equation}
While looking for solutions, we shall assume that the wave vector $\vec{q}$ of the field lies in $x$ direction, and that the field has a plane--wave expansion, i.e.
\begin{equation}
P(r_*, v, \vec{x})=\int \mathrm{d} \vec{q} ~e^{-i \omega t+i \vec{q}\cdot \vec{x}} ~\mathrm{p}(r_*, \omega, \vec{q}) \, 
\end{equation}
and solve the E.O.M. in the component form with the ansatz that all the $P$--form fields $P_{\mu_1 \mu_2 \cdots \mu_p}$ are zero except the longitudinal modes $P_{t \mu_2 \cdots \mu_p}$ and $P_{r_* \mu_2 \cdots \mu_p}$ \footnote{We can choose any particular set of $\mu_2 \cdots \mu_p$ coordinates such that the coordinate $x$ doesn't appear.} . The computations are similar to what has been carried out in section \ref{sec:2formfields} and \ref{sec:2formotoc5dim} with some differences which we have pointed out along the way. The variation of the action \eqref{eq:actionpform} leads to two independent equations of motion,
\begin{equation} \label{eq:beom1pform}
 \mathrm{p}_{r_*\mu_2\cdots\mu_p}(r_*) \left(q^2 f(r_*)-\omega^2\right)+i \omega~ \mathrm{p}_{t\mu_2\cdots\mu_p}'(r_*) = 0
\end{equation}
\begin{equation} \label{eq:beom2pform}
\omega ~\mathrm{p}_{t\mu_2\cdots\mu_p}(r_*) z(r_*)+i \left(-z(r_*)~\mathrm{p}_{r_*\mu_2\cdots\mu_p}'(r_*)+(d-2p)~\mathrm{p}_{r_*\mu_2\cdots\mu_p}(r_*) z'(r_*)\right)= 0 \,,
\end{equation}
where prime (${}^\prime$) denotes derivative with respect to $r_*$. Substituting $\mathrm{p}_{r_*\mu_2\cdots\mu_p}$ from \eqref{eq:beom1pform} in the second equation \eqref{eq:beom2pform} we get a second order differential equation in $\mathrm{p}_{t\mu_2\cdots\mu_p}$,
\small
\begin{equation} \label{eq:emainpform}
    \mathrm{p}_{t\mu_2\cdots\mu_p}''(r_*)- \partial_{r_*} \text{ln} \left[\left(\omega^2-q^2 f(r_*)\right)z(r_*)^{(d-2p)}\right]~ \mathrm{p}_{t\mu_2\cdots\mu_p}'(r_*) +  \left(\omega^2-q^2 f(r_*)\right) \mathrm{p}_{t\mu_2\cdots\mu_p}(r_*) = 0 \, .
\end{equation}
\normalsize
The near horizon analysis in the hydrodynamic limit yields the following expression,
\begin{align} \label{eq:ptyd2}
\mathrm{p}_{t\mu_2\cdots\mu_p}(r_*, \omega, q)=e^{-i \omega r_*} C(\omega, q) \Biggl[ 1 &+i \omega \int_{-\infty}^{r_*} d r_*\left(1-\left(\frac{z(r_{*})}{z_H}\right)^{(d-2p)}\right)  \\ \nonumber
& \left.+\frac{i q^2}{\omega} \int_{-\infty}^{r_*} d r_*\left(\frac{z(r_{*})}{z_H}\right)^{(d-2p)} f(r_{*})  
+\mathcal{O}\left(\lambda^2\right)\right] \, ,
\end{align}
and the normalization constant can be fixed by requiring $\displaystyle \lim _{r_* \rightarrow 0} \mathrm{p}_{t\mu_2\cdots\mu_p}(r_*, \omega, q) \rightarrow 1$ . The final expression for $\mathrm{p}_{t\mu_2\cdots\mu_p}$ takes the form,
\begin{equation}\label{eq:ptyd3}
\mathrm{p}_{t\mu_2\cdots\mu_p}(r_*, \omega, q)\sim e^{-i\omega r_*}\frac{1
+i  \omega H(r_*)
+i \frac{q^2}{(d-2p+1) \omega} z_{H} \left(1-\left(\frac{z(r_*)}{z_{H}}\right)^{(d-2p+1)}\right)}{1
+i\omega H(0)
+i \frac{q^2}{(d-2p+1) \omega} z_{H}}\, .
\end{equation}
We take note of the following points:
\begin{itemize}
    \item For $d=2p$ , $\displaystyle H(r_*)=\int_{-\infty}^{r_*} d r_*\left(1-\left(\frac{z(r_{*})}{z_H}\right)^{(d-2p)}\right)$ vanishes. Otherwise, following the discussion below \eqref{eq:btywithc}, under the $\omega \sim q^2$ scaling, $H(r_*)$ and $H(0)$ can be neglected.
    \item The expression \eqref{eq:ptyd3} is valid for $d>(2p-1)$, in absence of any boundary divergences. In all such case the diffusion constant $\displaystyle D=\frac{z_{H}}{d-2p+1}=\frac{d+1}{4 \pi(d-2p+1) T}$ .
    \item Logarithmic divergences are observed when $d=2p-1$. In this case, equation \eqref{eq:ptyd2} can be regularized following section \ref{logdivergencebfield} and the new expression replacing \eqref{eq:ptyd3} is similar to \eqref{eq:btyfinal2} where the diffusion constant $\displaystyle D\sim z_H \ln{\left(\frac{z_H}{\bar{z}}\right)}=\frac{2p}{ 4 \pi T} \ln{\left(\frac{2p}{4 \pi\bar{z}T}\right)}$ and $\displaystyle \bar{z}=z_\Lambda e^{-\frac{1}{\kappa(\Lambda)}}$ is the RG--invariant scale. 
    \item For $d<(2p-1)$, there are power law divergences of the form $\displaystyle 1/r_*$, $\displaystyle 1/r_*^2$ and so on. Again, the boundary divergences can be regularized by resorting to holographic renormalization, similar to the discussion in section \ref{logdivergencebfield}. 
    %A specific example of a 3--form field in five dimensions is discussed below. Similar analysis can be performed for other form fields in other dimensions and the general expression for a $p$--form is \eqref{eq:ctyfinal1pform}.%
% For a $p$--form field propagating in $d+2$ dimensional asymptotic AdS geometry such that $d<(2p-1)$, the associated boundary divergences can be regularized in a similar fashion. %
In these cases, the final expression for the $p$--form field takes the form \footnote{By a slight abuse of notation, $z$ here is now the renormalized $z$.},
\begin{equation} \label{eq:ctyfinal1pform}
   \mathrm{p}_{tyw\dots} (r_*, \omega, q)=e^{-i \omega r_*} \left[ \frac{1+ \frac{i q^2}{(d-2p+1)\omega} z_H \left(1-\left(\frac{z_H}{z}\right)^{-(d-2p+1)}\right) }{1+\frac{i q^2}{(d-2p+1)\omega} z_H\left(1-\left(\frac{z_H}{z_{RG}}\right)^{-(d-2p+1)}\right)}+\mathcal{O}\left(\lambda^2\right) \right] \,,
\end{equation}
where $z_{RG}$ is the RG--invariant scale and it is defined by the following relation,
\begin{equation}
z_{RG}^{(d-2p+1)}= z_\Lambda^{(d-2p+1)} -{\frac{(d-2p+1)}{\kappa(\Lambda)}} \, .
\end{equation}
The diffusion constant in this case is defined as, 
\begin{equation}
\begin{aligned}
    D_p&=\frac{z_H}{(d-2p+1)}\left(1-\left(\frac{z_{H}}{z_{RG}}\right)^{-(d-2p+1)}\right)\\
    &=\frac{d+1}{4 \pi(d-2p+1)T}\left(1-\left(\frac{d+1}{4 \pi z_{RG}T}\right)^{-(d-2p+1)}\right)\, .
\end{aligned}
\end{equation}
%The corresponding $\mathrm{OTOC}$ has the following expression,
%\begin{equation} \label{eq:otocpfivedim}
%\mathrm{OTOC}=\left(P_1, P_2\right) \sim \frac{1}{ D_p^{\frac{d}{2}+1}\left[\ln \left(2+\frac{\Delta_O}{N} e^{\frac{2 \pi}{\beta} t_{12}-\mu\left|\vec{x}_{12}\right|}\right)\right]^{\frac{d}{2}}}\,  ,
%\end{equation}

As an example, the 3--form field case is worked out in detail below.

    \item The inner product between two $p$--form fields get a factor $(g^{yy})^{p}$. For example, in the 2--form field case there are two indices in the $B_{\mu\nu}$, so there is a $(g^{yy})^2$ factor in the numerator. While this does affect the $\mathrm{OTOC}$ computations, %with or without the shockwave% 
    it doesn't affect the late time behavior of $\mathrm{OTOC}$.
    \item Assuming the $SO(d-1)$ spherical symmetry of the solutions, the late time power law tail of the $\mathrm{OTOC}$ remains the same for all form fields and only depends on the dimensions of the spacetime,   $\displaystyle \mathrm{OTOC} \sim t^{-\frac{d}{2}}$, where $d$ is the number of transverse directions. However, we note that the higher form symmetry currents have extra spatial indices which in principle curtails the spherical symmetry. In the earlier computations for the 2--form field, we considered the boundary current operator as a point operator with a spherical symmetry along the $d$--dimensional spatial plane in the boundary but the current operator $J_{ty}$ has a spatial index that breaks the spherical symmetry in the $y$--direction. For this reason, the real space form of the bulk field solution in \eqref{eq:btyfinal2} will necessarily depend on the transverse directions $\vec{x}$ that are also perpendicular to $y$. Effectively, all the computations done in the main sections of the paper will go through (the boundary behavior and the diffusion constant remains the  same) with the restriction that now the $d$--dimensions effectively becomes $(d-1)$--dimensions. In section \ref{sec:b5dim}, restricting to $2=(3-1)$ spatial dimensions in the boundary ($d=3$ for five--dimensional bulk), we should ideally consider $\mathrm{OTOC}$s with the line operator $J^s(t, \vec{x}):=\int_{-\infty}^{+\infty} d y J_{t y}(t, y, \vec{x})$ (which effectively lives in $(d-1)$--dimensions at the boundary) instead of a point operator $J_{ty}$. The choice of line operators over point operators changes the late time power law behavior from $\sim t^{-\frac{3}{2}}$ to $\sim t^{-1}$. Similarly, for a $p$--form field, the corresponding current operator $J_{t{\mu}_2 \dots {\mu}_p}$ breaks the spherical symmetry in ${\mu}_2$ to ${\mu}_p$ directions. Therefore, the late time behavior for the $\mathrm{OTOC}$ involving the $(p-1)$--form symmetry currents reads, $\sim t^{-\frac{d-(p-1)}{2}}$ instead of $\sim t^{-\frac{d}{2}}$. We conclude that whether we take full spherical symmetry of the solutions or reduce it appropriately, the usual late time exponential decay of the $\mathrm{OTOC}$ is replaced by a power law decay in the presence of a conservation law.
    %It seems to be a universal feature of the form fields.
\end{itemize}

\subsubsection*{A 3--form field $C_{\mu \nu \rho}$ in five dimensional AdS--Schwarzschild geometry} \label{3form5d}
%The field strength is  
%\begin{equation}
%F^{(4)}_{\mu_1 \mu_2 \mu_3 \mu_4}=\partial_{\mu_1} C_{\mu_2 \mu_3 \mu_4}-\partial_{\mu_2} C_{\mu_3 \mu_4 \mu_1}+\partial_{\mu_3} C_{\mu_4 \mu_1 \mu_2}-\partial_{\mu_4} C_{\mu_1 \mu_2 \mu_3}.
%\end{equation}
A near horizon analysis, as discussed in previous sections gives the following expression for the $c_{tyw}$ field,
\begin{align} \label{eq:ctyz5}
c_{tyw}(r_*, \omega, q)=e^{-i \omega r_*} C(\omega, q) \Biggl[ 1 &+i \omega \int_{-\infty}^{r_*} d r_*\left(1-\left(\frac{z_H}{z(r_{*})}\right)^3\right)  \\ \nonumber
& \left.+\frac{i q^2}{\omega} \int_{-\infty}^{r_*} d r_*\left(\frac{z_H}{z(r_{*})}\right)^3 f(r_{*})  
+\mathcal{O}\left(\lambda^2\right)\right] \, ,
\end{align}
where, as before, the second term can be neglected. Performing the integral, the expression for $c_{tyw}$ can be written as,
\begin{align} \label{eq:ctyz5_1}
c_{tyw}(r_*, \omega, q)=e^{-i \omega r_*} C(\omega, q) \left[1-\frac{i q^2}{\omega}\frac{z_H}{2}\left(1-\left(\frac{z_H}{z}\right)^2\right)\right] \, ,
\end{align}
where we encounter the $1/z^2$ divergence at the boundary.
This boundary divergence can be regularized by a double trace deformation counterterm as discussed in section \ref{logdivergencebfield}. It induces a shift in the boundary solution for the $C$--field,
\begin{equation}
\frac{\partial S}{\partial J_{\alpha \beta \sigma}} = \frac{\gamma^2}{2 \kappa(\Lambda)} \frac{\partial {J_{\mu \nu \rho} J^{\mu \nu \rho} }}{\partial J_{\alpha \beta \sigma}}    =   \frac{\gamma^2}{ \kappa(\Lambda)} J^{\alpha \beta \sigma}  \, .
\end{equation}

For us only the component $C_{tyw}$ is relevant and its boundary value gets a shift, $ \displaystyle \partial S/\partial J_{tyw}=  (\gamma^2/\kappa(\Lambda)) J^{tyw}$.
%On the other hand, using \eqref{eq:Jdef} and 
%\begin{equation}
%    J^{\mu \nu}=-\frac{1}{\gamma^2}\sqrt{-g} H^{r_* \mu \nu},
%\end{equation}
%$H_{r_{*}ty}=\partial_{r_*}B_{ty}-\partial_{t}B_{r_{*}y}$.
%substituting $b_{r_{*}y}$ from \eqref{eq:beom1}, we can write $H_{r_{*}ty}$ as,
The relevant component of the field strength $F^{(4)}=d C^{(3)}$ can be written in terms of $c_{tyw}$ as $ \displaystyle F^{(4)}_{r_{*}tyw}=\frac{q^2 f} {(q^2 f -\omega^2)}c_{tyw}^{\prime}$,
which leads to  $ \displaystyle J^{tyw}=\frac{1}{\gamma^2}C(\omega,q) z_{H}^3\frac{i q^2}{\omega}$ in the hydrodynamic limit.
%\begin{equation}
%\begin{aligned}
%    J^{tyw} %&=-\frac{1}{\gamma^2}\sqrt{-g} g^{r_{*}r_{*}} g^{tt} g^{yy} H_{r_{*} t y}\\&
%    =\frac{1}{\gamma^2} \frac{q^2 f} {(q^2 f -\omega^2)} z^3 ~ c_{tyw}^\prime \, .
%    \end{aligned}
%\end{equation}
%In the hydrodynamic small $\omega$ limit, we can write $J^{ty}$ as, 
Incorporating the shift, the near boundary expression for $c_{tyw}$ is as follows,
\begin{equation}
\begin{aligned}
    c_{tyw}(r_*\to 0)&=C(\omega,q) \left[1- \frac{i q^2}{\omega} \frac{z_H}{2}\left(1-\left(\frac{z_H}{z}\right)^2\right) +\frac{1}{\kappa(\Lambda)}\frac{i q^2}{\omega} z_H^3\right]\\&
    %=C(\omega,q) \left[1+ \frac{i q^2}{\omega} z_H \text{ln}z_H - \frac{i q^2}{\omega} z_H \text{ln}\bar{z}\right]\\&
    =C(\omega,q) \left[1- \frac{i q^2}{2\omega} z_H \left(1-\left(\frac{z_H}{z_*}\right)^2\right)\right] \,,
\end{aligned}
\end{equation}
where $z_*$, defined as $\displaystyle \frac{1}{z_*^2}= \frac{1}{z_\Lambda^2} +{\frac{2}{\kappa(\Lambda)}}$, is the RG--invariant scale \footnote{For a perspective involving the cut--off independence of a physical source leading up to a fixed point equation of a $\beta$--function, followed by a renormalization group (RG) equation for the running double--trace coupling $\kappa(\Lambda)$, see \cite{Grozdanov_2018}.}. As discussed before, the  normalization constant $C(\omega,q)$ can be fixed by demanding $\lim_{r_*\to 0} c_{tyw}=1$, leading up to the final expression for $c_{tyw}$ as \footnote{By a slight abuse of notation, $z$ here is the renormalized $z$.},
\begin{equation} \label{eq:ctyfinal1}
   c_{tyw} (r_*, \omega, q)=e^{-i \omega r_*} \left[ \frac{1- \frac{i q^2}{2\omega} z_H \left(1-\left(\frac{z_H}{z}\right)^2\right) }{1-\frac{i q^2}{2\omega} z_H\left(1-\left(\frac{z_H}{z_*}\right)^2\right)}+\mathcal{O}\left(\lambda^2\right) \right] \,.
\end{equation}
The diffusion constant in this case is defined as,
\begin{equation}
    D_3=-\frac{z_H}{2}\left(1-\left(\frac{z_H}{z_*}\right)^2\right)=-\frac{1}{2\pi T}\left(1-\left(\frac{1}{\pi z_* T}\right)^2\right)\, .
\end{equation}
The $\mathrm{OTOC}$ can be computed in a similar fashion as discussed in section \ref{sec:b5dim} and is given as,
\begin{equation} \label{eq:otocC3fivedim}
\mathrm{OTOC}=\left(C_1, C_2\right) \sim \frac{1}{ D_3^{\frac{3}{2}+1}\left[\ln \left(2+\frac{\Delta_O}{N} e^{\frac{2 \pi}{\beta} t_{12}-\mu\left|\vec{x}_{12}\right|}\right)\right]^{\frac{3}{2}}}\,  ,
\end{equation}
which exhibits the same power law decay, $\mathrm{OTOC}$ $\sim t^{-\frac{3}{2}}$ at late times \footnote{If instead of point current operator we choose the extended current operator $J^s(t, \vec{x}):=\iint d y d w J_{t y w}(t, y, w, \vec{x})$, the power law changes to $\sim t^{-\frac{1}{2}}$.}.

\section{Conclusions}
This work is primarily motivated by an interesting recent observation regarding the late time behavior of the $\mathrm{OTOC}$ involving conserved current operators which shows a diffusive power law tail instead of the usual exponential decay. We have extended the analysis by including the conserved currents that are associated to higher--form global symmetry at the boundary.  
%In this paper we studied the late time behaviour of $\mathrm{OTOC}$ between conserved current operators associated with higher form $U(1)$ global symmetry and heavy scalar operators as a diagnostic of quantum chaos. In the absence of a local consevation law, $\mathrm{OTOC}$ decays to zero exponentially fast. However, in the presence of a conserved current, the decaying process becomes diffusive; a power law tail is observed.
In this paper, we focused on the 2--form antisymmetric $B$--fields propagating in five dimensional AdS--Schwarzschild geometry and its solutions in the near horizon region. By considering only the in--falling modes and invoking the hydrodynamic limit, we have solved the relevant equations of motion for the vector modes after regularizing the logarithmic divergence at the AdS boundary. The scrambling property survives the regularization and the  $\mathrm{OTOC}$ has been computed as the inner product between asymptotic `in' and `out' states, which in this case becomes equivalent to computing the inner product between two $2$--form fields with or without the shockwave resulting  from the backreaction of the heavy scalar operators on the unperturbed AdS--Schwarzschild geometry. The presence of the shockwave is essentially captured by a shift in the outgoing Kruskal coordinate $v$, proportional to the energy carried by quanta of the very heavy scalar operator. The $\mathrm{OTOC}$ between the conserved $U(1)$ current operators and the heavy scalar operator displays a diffusive power law tail at late times. In the later half of the paper, we have generalized the case of 2--form fields to $p$--forms in $(d+2)$ dimensional AdS--Schwarzschild geometry. However, it is observed that the late time behavior of $\mathrm{OTOC}$ with a $U(1)$ charge conservation law always decays with a power law tail.
%, $\mathrm{OTOC} \sim t^{-\frac{d}{2}}$ where $d$ is the number of transverse directions. 
%It seems to be a universal feature of the form fields with $U(1)$ charge conservation. 
Due to the hydrodynamical property of the conserved current, the particles sourced by these conserved operators in the bulk spreads over a large region of the spacetime leading up to this kind of diffusive behavior at late times.

In the present work, we have evaluated the wave functions for the form fields to leading order in $\omega$ and $q$ in the hydrodynamic limit. It is possible to include higher order terms in $\lambda$ and perform a similar computation. The dispersion relation will get corrected to higher orders in $q$ and the wavefunctions too will pick up higher order corrections. We expect the power law tails for late time $\mathrm{OTOC}$s to survive the higher order corrections but still it  would be interesting to compute the  $\mathrm{OTOC}$s with higher precision. It is also possible to include loop corrections due to graviton and calculate the $\mathrm{OTOC}$s with the modified wave functions. In our computations, we have introduced the bulk form field in the probe approximation and neglected its backreaction to the geometry. Though challenging,  it might be possible to move away from the probe approximation. Another interesting future direction would be to investigate the dynamics, diffusion constants and $\mathrm{OTOC}$s for non--abelian conserved charges, instead of the $U(1)$ charge studied in this paper. It will also be interesting to repeat the same analysis for Kerr--AdS geometries in arbitrary number of dimensions \cite{Lunin:2017drx, Lunin:2019pwz}.

\subsection*{Acknowledgments}
A.C. acknowledges the Strings 2024 held at CERN, Geneva, String--Math 2024 and School \& Workshop on Number Theory and Physics held at ICTP,  Trieste for their hospitality and creating stimulating environments where part of the work was completed. A.C. also acknowledges the NSM 2024 held at IIT Ropar where the work was presented leading to useful discussions. The work of A.C. is supported by IIT Bhubaneswar Seed Grant SP–103. K.S. thanks Department of Physics, New Alipore College. S.M. thanks ICTS, Bangalore for warm hospitality and the String Theory and Quantum Gravity group for the discussions while presenting the work in an ICTS string seminar. The work of S.M. is supported by fellowship from CSIR, Govt. of India.

\begin{appendix}
\section{$B$--field in seven dimensional AdS--Schwarzschild geometry}\label{sec:b7dim}
In this appendix, we study the $B$--field as discussed in \ref{sec:2formfields}, but without any boundary divergence.
%In this section case we study the late time behaviour of $OTOC$ with two conserved current operators and the heavy scalar operators for the same $B$-field but now for 7-dimensional AdS-Schwarzschild black holes.
%\subsubsection*{E.O.M. and its solutions}
As before, we shall focus on the vector modes and consider the ansatz where all $B_{\mu \nu}$ fields are zero except $B_{ty}$ and $B_{r_*y}$. For seven dimensional AdS--Schwarzschild geometry,
 $\displaystyle f(r_{*})=1-\left(\frac{z(r_{*})}{z_{H}}\right)^{6}$ and 
%From the two independent equations \eqref{eq:beom1} and \eqref{eq:beom2}, we get a second order differential equation in $b_{ty}$, equation \eqref{eq:emain} for $\displaystyle f(r_{*})=1-\left(\frac{z(r_{*})}{z_{H}}\right)^{6}$.%
the set of E.O.M. reduces to two independent equations: 
\begin{equation} \label{eqq1}
 b_{r_*y}(r_*) \left(q^2 f(r_*)-\omega^2\right)+i \omega b_{ty}'(r_*) = 0
\end{equation}
\begin{equation} \label{eqq2}
\omega b_{ty}(r_*) z(r_*)-i \left(z(r_*) b_{r_*y}'(r_*)-b_{r_*y}(r_*) z'(r_*)\right)= 0
\end{equation}
%Here prime denote derivative w.r.t. $r_*$. Now we substitute $b_{r_*y}$ from \eqref{eqq1} in the second equation \eqref{eqq2},
By eliminating $b_{r_* y}$, we get a second order differential equation in $b_{ty}$,
%\begin{equation}
%    b_{ty}''(r_*)+  \left(\frac{q^2 f'(r_*)}{\omega^2-q^2 f(r_*)} -\frac{z'(r_*)}{z(r_*)}\right) b_{ty}'(r_*) +  \left(\omega^2-q^2 f(r_*)\right) b_{ty}(r_*) = 0\, ,
%\end{equation}
%which can be rewritten as,
\begin{equation} \label{emainqq}
    b_{ty}''(r_*)- \partial_{r_*} \text{ln} \left[z(r_*)\left(\omega^2-q^2 f(r_*)\right)\right]~ b_{ty}'(r_*) +  \left(\omega^2-q^2 f(r_*)\right) b_{ty}(r_*) = 0 \, .
\end{equation}
The above equation has two independent solutions, wherein as $r_* \rightarrow \infty$, they behave like $e^{ \pm i \omega r_*}$, corresponding respectively to the out--going and in--falling boundary conditions at the horizon. As before, we focus on the in--falling mode, and the out--going mode if required can be obtained by complex conjugation. Analogous to \eqref{eq:btywithc}, the solution for $b_{ty}$ can be found explicitly in the hydrodynamic limit upto to $\mathcal{O}(\lambda)$,
\begin{align} \label{eq:bty7}
b_{ty}(r, \omega, q)=e^{-i \omega r_*} C(\omega, q) \Biggl[ 1 &+i \omega \int_{-\infty}^{r_*} d r_*\left(1-\left(\frac{z(r_{*})}{z_{H}}\right)\right)  \\ \nonumber
& \left.+\frac{i q^2}{\omega} \int_{-\infty}^{r_*} d r_*\left(\frac{z(r_{*})}{z_{H}}\right) f(r_{*})  
+\mathcal{O}\left(\lambda^2\right)\right] \, .
\end{align}

The above equation has no logarithmic divergences and the normalization constant $C(\omega,q)$ can be straightforwardly fixed by requiring $\displaystyle \lim _{r_* \rightarrow 0} b_{ty}(r_*, \omega, q) \rightarrow 1$ . Again, the second term can be neglected (see the discussion below equation \eqref{eq:btywithc}) and we arrive at the final expression for $b_{ty}$ ,
\begin{equation}
b_{ty}(r_*, \omega, q)=e^{-i\omega r_*}\left[\frac{1+ %i \omega H(r_*)+
i \frac{q^2}{2 \omega} z_{H} \left(1-\left(\frac{z(r_*)}{z_{H}}\right)^2\right)}{1+%i \omega H(0)+
i \frac{q^2}{2 \omega} z_{H}}+\mathcal{O}\left(\lambda^2\right)\right]\, ,
\end{equation}
where the factor of $\displaystyle z_{H}/2$ can be identified as the diffusion constant $D$. Following a similar discussion as in section \ref{sec:b5dim}, by considering  the hydrodynamic limit and $\omega$ scaling as $\omega \sim q^2$, the near horizon i.e. $z \sim z_{H}$ expression for the $b_{ty}$ has the following form, 
\begin{equation}
b_{ty}(r_*, \omega, q) \sim e^{-i \omega r_*} \frac{\omega}{\omega+i D q^2} \, .
\end{equation}
The above expression is the same as of a 1--form gauge field $A_\mu$ , in five dimensional  AdS--Schwarzschild geometry as discussed in \cite{Cheng:2021mop} but with a different value for $D$. As we have seen in section \ref{sec:2formotoc5dim}, where we computed the $\mathrm{OTOC}$s in the late time regime, the particular value of $D$ is of little consequence. 
%One important point to note here that there will be no divergences in this case unlike in 5 dimensions. 

\subsection*{$B$--field in six dimensions}
For completeness we present here the results for the $B$--field in six spacetime dimensions with $\displaystyle f(r_{*})=1-\left(\frac{z(r_{*})}{z_{H}}\right)^{5}$ for the AdS--Schwarzschild geometry. As before, we consider only the longitudinal mode with the ansatz, all $B_{\mu \nu}$ are zero except $B_{ty}$ and $B_{r_*y}$ . From \eqref{eombmn}, the E.O.M. for $b_{t y}$ is a second order differential equation,
\begin{equation} \label{eq:emain6d}
    b_{ty}''(r_*)- \partial_{r_*} \text{ln} \left(\omega^2-q^2 f(r_*)\right)~ b_{ty}'(r_*) +  \left(\omega^2-q^2 f(r_*)\right) b_{ty}(r_*) = 0 \, .
\end{equation}
Considering only the in--falling modes in the hydrodynamic limit with $\omega \sim q^2$ scaling,  the resulting near horizon analysis gives the following expression for $b_{ty}$ ,
\begin{align} \label{eq:bty6d}
b_{ty}(r, \omega, q)=e^{-i \omega r_*} C(\omega, q) \left[ 1 +\frac{i q^2}{\omega} \int_{-\infty}^{r_*} d r_* f(r_{*})  
+\mathcal{O}\left(\lambda^2\right)\right] \, ,
\end{align}
where the normalization constant $C(\omega,q)$ can be fixed by requiring that $\displaystyle \lim _{r_* \rightarrow 0} b_{ty}(r_*, \omega, q) \rightarrow 1$. We further note that unlike the analogous expressions in five and seven dimensions the term involving the $H(r_*)$ is absent in \ref{eq:bty6d} . As there is no divergence, the final expression for $b_{ty}$ with the diffusion constant, $ 
\displaystyle D=  \int_{-\infty}^{0} d r_*  f(r_*) =z_H$ can be written as 
%\begin{equation}\label{btyfinal6d}
%b_{ty}(r_*, \omega, q)=e^{-i\omega r_*}\left[\frac{1+
%i \frac{q^2}{ \omega} z }{1+
%i \frac{q^2}{ \omega} z_{H}}+O\left(\lambda^2\right)\right]\, ,
%\end{equation}
\begin{equation}\label{btyfinal6d}
b_{ty}(r_*, \omega, q) \sim e^{-i\omega r_*}\frac{1+
i \frac{q^2}{ \omega} z }{1+
i \frac{q^2}{ \omega} D} \, .
\end{equation}

\end{appendix}

\bibliographystyle{unsrt} 
\bibliography{OTOC_higherform.bib}

\begin{comment}

\end{comment}
\end{document}